\documentclass[aps,twocolumn,superscriptaddress,floatfix,longbibliography]{revtex4-1}
\usepackage{tikz}
\usepackage{lipsum}
\usepackage{graphicx}
\usepackage[export]{adjustbox}
\usepackage{amsmath,amssymb,wasysym}
\usepackage{braket}
\usepackage{mathtools}
\usepackage{mathrsfs}
\usepackage{verbatim}
\usepackage{makecell}
\usepackage{cases}
\usepackage[compat=1.0.0]{tikz-feynman}
\usepackage{hyperref}
\usepackage{xcite}
\usepackage{xr}

\usepackage{tikz}
\usepackage{xcolor}
\usetikzlibrary{shapes.misc}
\usetikzlibrary{patterns}
\usetikzlibrary{calc,intersections,decorations.pathmorphing,decorations.markings,decorations.pathreplacing}
\usetikzlibrary{shadows}
\usetikzlibrary{fadings}

\usepackage{pgfplots}
\makeatletter
\newcommand*{\addFileDependency}[1]{
	\typeout{(#1)}
	\@addtofilelist{#1}
	\IfFileExists{#1}{}{\typeout{No file #1.}}
}
\makeatother
\newcommand*{\myexternaldocument}[1]{%
	\externaldocument{#1}%
	\addFileDependency{#1.tex}%
	\addFileDependency{#1.aux}%
}
\newcommand{\im}{\mathbf{i}}
\newcommand{\R}{\mathbb{R}}

\newcommand{\lmax}{\lambda_{\max}}

\newcommand{\sgn}{\mathrm{sign}}
\DeclareMathOperator{\Tr}{Tr}

\newcommand{\BigO}{\mathcal{O}}
\newcommand{\smallo}{\mathrm{o}}

\def\pbraket#1{\mathinner{\left(#1\right)}}

\myexternaldocument{supp}

\begin{document}
	\title{Low-rank Sachdev-Ye-Kitaev models}	
	\author{Jaewon Kim}
	\affiliation{Department of Physics, University of California, Berkeley, CA 94720, USA}
	\author{Xiangyu Cao}
	\affiliation{Department of Physics, University of California, Berkeley, CA 94720, USA}
	\author{Ehud Altman}
	\affiliation{Department of Physics, University of California, Berkeley, CA 94720, USA}
	\date{\today}
	\begin{abstract}
		Motivated by recent proposals of experimental realization of fast scramblers, we study a family of solvable variants of the ($q=4$) Sachdev-Ye-Kitaev model in which the rank and eigenvalue distribution of the coupling matrix $J_{ij,kl}$ are tuneable. When the rank is proportional to the number of fermions, the low temperature behavior is sensitive to the eigenvalue distribution. We obtain a complete classification of the possible non-Fermi liquid quantum phases. These include two previously studied phases whose fermion scaling dimension depends continuously on the rank; we show that they are maximally chaotic, but necessitate {an extensively degenerate or negative semidefinite coupling matrix}. More generic distributions give rise to ``almost Fermi liquids'' with a scaling dimension $\Delta = 1/2$, but which differ from a genuine Fermi-liquid in quasi-particle decay rate, quantum Lyapunov exponent and/or specific heat.
	\end{abstract}
	\maketitle
	
	\section{Introduction} 
	The Sachdev-Ye-Kitaev~\cite{Sachdev:1992fk,Kitaev:2015} model, in its simplest form, describes a large number of Majorana fermions with all-to-all random interactions:
	\begin{equation}
		H = \sum_{ijkl=1}^N J_{ij,kl} \gamma_i \gamma_j \gamma_k \gamma_l  \,.  \label{eq:SYKintro}
	\end{equation}
	At low temperatures, this exactly solvable model describes a peculiar non Fermi liquid which has a large symmetry, and a quantum Lyapunov exponent that saturates the universal bound on chaos~\cite{maldacena2016bound}. These features made it an attractive platform to study a wide range of topics, e.g., strongly correlated electrons, many-body quantum chaos, and black hole information scrambling, each generating a flurry of recent activities~\cite{Sachdev:1992fk, Kitaev:2015,Maldacena:2016hyu,Kitaev:2017awl,Jevicki:2016bwu,Bagrets:2016cdf,Banerjee:2016ncu,Sachdev:2015efa,Bi:2017yvx,2017PhRvL.119u6601S,Patel:2018zpy,Kolovsky:2016irf,Scaffidi:2017ghs,Gu:2016oyy,Gao:2016bin,Maldacena:2018lmt,Kim:2019upg,parker2018universal,Lian:2019axs}. 
	
	Historically, the SYK model originated from the Sachdev-Ye (SY) model of quantum random spin magnet \cite{Sachdev:1992fk}:
	\begin{equation}
		H = \frac{1}{\sqrt{NM}}\sum_{a,b=1}^{N} U_{ab} \mathbf{S}_a\cdot\mathbf{S}_b,
		\label{eq:SYintro}
	\end{equation}
	where $\mathbf{S}_a$ are some $SU(M)$ spin operators. The SYK Hamiltonian was conceived by Kitaev as a variant of the fermionic representation of \eqref{eq:SYKintro} in the double scaling limit $M,N \to \infty$: schematically, a spin operator is represented by a fermion bilinear, and the coupling matrix $U_{ab}$ by $J_{ij,kl}$. Although the SY model beyond the double-scaling limit is not exactly solvable, it is more amenable to experimental realization. In particular, coupling cold atom ensembles to optical cavity modes provides a promising way of generating the all-to-all interaction between atomic spins~\cite{PhysRevLett.91.203001,2007Natur.449..443M,PhysRevLett.104.073602,2013Sci...342.1494V,Kollar2017,Leonard2017,norcia2018cavity,Marino2018,strack2011dicke,swingle2016measuring,2019arXiv190410966B}. In these platforms, the rank of the matrix $U_{ab}$ is controlled by the number of coupled cavity modes, which is usually rather small. The effect of having a low-rank matrix has been studied in detail in \cite{2019arXiv190410966B}, where it was shown that the resulting quantum dynamics is integrable even at infinite temperature. These findings leave one wondering how large a rank is necessary to access SYK physics. This question is further complicated by the double scaling limit: in the standard SYK model \eqref{eq:SYKintro}, $J_{ij,kl}$ has independent coefficients and is a matrix of super-extensive rank $\propto N^2$, whereas in the SY model~\eqref{eq:SYintro} with a fixed $M$, $U_{ab}$ has an extensive rank $\propto N$. Therefore, a solvable variant of the SYK model where $J_{ij,kl}$ has tuneable rank should be beneficial to better understanding random quantum magnets beyond large $M$.
	
	\begin{figure}
		\begin{center}
			\includegraphics[width=0.75\columnwidth]{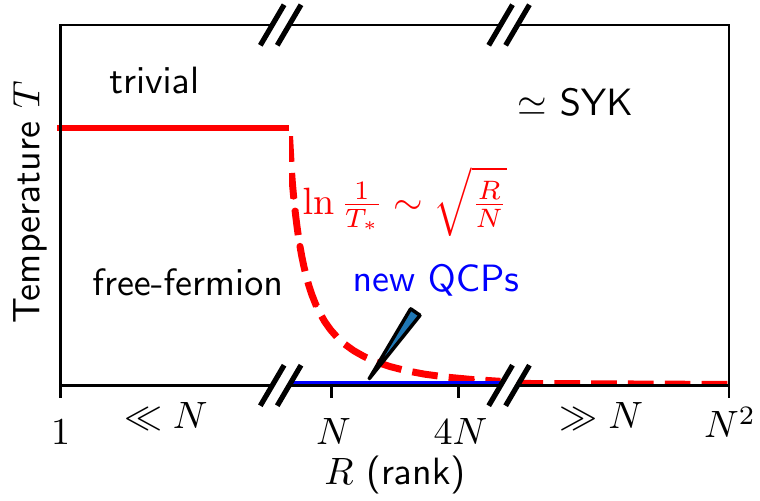}
		\end{center}
		\caption{A qualitative phase diagram (main plot) and a sketch (inset) of the low-rank SYK model. $N$ Majorana fermions (blue dots) are coupled by random all-to-all 4-body interactions, mediated by $R$ boson modes; $R$ is also the rank of the coupling matrix \eqref{eq:Jijkl}. The model is non-interacting when $R \ll N$ and equivalent to SYK when $ R \gg N$. When $R \propto N$, the IR fixed point, governing $T \lesssim T_*$, depends on the eigenvalue distribution of the coupling matrix, see Table~\ref{table:ExtensiveCase}.}
		\label{fig:phasediagram}
	\end{figure}
	Such a model has recently  been considered by several authors in different contexts: for example, to showcase the instability of the SYK fixed point towards a Fermi-liquid phase~\cite{Bi:2017yvx}, and to model Cooper pairing in non-Fermi liquids ~\cite{wang19,Esterlis:2019ola}. In the latter context, the rank equals the number of phonon modes coupled to the electrons. So far, it has been understood that the extensive rank ($R\sim N$) regime is the most interesting, whereas $R \gg N$ leads back to the standard SYK model and $R \ll N$ to a non-interacting model~\cite{masaki16,Bi:2017yvx}, see Fig.~\ref{fig:phasediagram}. 
	
	What was overlooked, however, is the role of the {eigenvalue distribution} of the coupling matrix, or equivalently, the distribution of fermion-boson/spin-boson couplings. In this paper, we fill in this gap by solving a family of ``low-rank SYK models'' where $J_{ij,kl}$ has a tuneable eigen-distribution. Our main contribution is an essentially complete classification table (Table~\ref{table:ExtensiveCase}) of four universality classes of distributions, which give rise to distinct gapless quantum phases. Among them, previous works~\cite{Bi:2017yvx,wang19,Esterlis:2019ola} studied two classes (III and IV in our classification), which we show are indeed SYK-like fast scramblers with extensive residual entropy. The new classes (I and II), corresponding to more generic distributions, exemplify quantum phases that are almost, but not quite, Fermi liquids. 
	
	As an application, we revisit the proposal put forward in Ref.~\cite{2018PhRvL.121c6403C} of realizing the SYK$_4$ model with electrons in the zeroth Landau level of a graphene flake with irregular boundaries. In this system the random four fermion interactions arise from the coulomb interactions, projected on to the zeroth Landau level. We argue that the model realized is actually a low-rank SYK, with extensive rank and of Class IV. 
	
	The rest of the paper is organized as follows. Sec.~\ref{sec:model} defines the model and provides some preliminary discussions, including the Schwinger-Dyson equations.  Sec.~\ref{sec:supersub} discuss the low and high rank limits of the sub-extensive and super-extensive rank regimes. Sec.~\ref{sec:results} mark the start of our discussion of the extensive rank regime: the four universality classes are introduced, and the correlation function of each class is studied. Sec.~\ref{sec:thermodynamics} presents both analytical and numerical results on the low temperature thermodynamics of the model at extensive ranks. Sec.~\ref{sec:chaos} studies the quantum chaos of the model at the extensive rank regime. Sec.~\ref{sec:experiments} discusses the graphene cornflak proposal~\cite{2018PhRvL.121c6403C}. We conclude in Sec.~\ref{sec:discussion}.

	\section{Model \& Schwinger-Dyson Equations}\label{sec:model}
	The Hamiltonian of the low-rank SYK model has the same form as~\eqref{eq:SYKintro}, but the coupling constants form a rank $R$ matrix: 
	\begin{align}
		J_{ij,kl} = \frac12 \sum_{n=1}^{R} \lambda_n  u_{ij}^{(n)}  u_{kl}^{(n)} \quad R = \gamma N + \smallo(N) \,.
		\label{eq:Jijkl}
	\end{align}
	Above, $\gamma = R/N$ is the rescaled rank and we shall mostly focus on the extensive rank regime where $\gamma = \BigO(1)$. $\{ u_{ij}^{(n)} \}$ are independent Gaussian random variables with zero mean and satisfying
	\begin{equation}
		\overline{ u_{ij}^{(n)} u_{kl}^{(m)} } = \frac1{N^2} \delta_{ik}\delta_{jl} \delta_{nm} \,. \label{eq:uij_coupling}
	\end{equation}
	
	Finally, we assume the eigenvalues $\{\lambda_n \}$~\footnote{It is justified to call $\lambda_n$ eigenvalues because in the large-$N$ limit, $\left\{u_{ij}^{(n)}\right\}$ has the same law as $R$ orthogonal random vectors in $\R^{N^2}$, so \eqref{eq:Jijkl} is effectively an eigen-decomposition.} to have a well-defined distribution 
	\begin{equation}
		\rho(\lambda) := \frac{1}{R} \sum_{n=1}^R \delta(\lambda - \lambda_n)  \label{eq:rholambda}
	\end{equation}
	in the $N\to \infty$ limit, such that  $\lmax := \max_n \lambda_n$ is also the right edge of $\rho$'s support.
	
	The above model is solvable for any $\rho(\lambda)$ in the large-$N$ limit, by essentially the same Hubbard-Stratonovich (HS) decoupling method used in Ref~\cite{Bi:2017yvx}. Indeed, the Hamiltonian can be rewritten as follows 
	\begin{equation}
		H = -\sum_{n=1}^R \frac12 \lambda_n Q_n^2 \text{ where }  Q_n := \sum_{i,j=1}^N  \im u_{ij}^{(n)}\gamma_i \gamma_j \label{eq:Qndef}
	\end{equation} 
	are a set of random fermion bilinears. The HS transformation then introduces the bosons $\{\phi_n\}_{n=1}^R$, for decoupling each of the $Q_n^2$ terms respectively. This results in the following Lagrangian:
	\begin{equation}
		\mathcal{L} =  \sum_j  \gamma_j \dot{\gamma_j} + \sum_n \left( \lambda_n^{\frac12} \phi_n Q_n + \frac{\phi_n^2}2 \right) \,. 
		\label{eq:H}
	\end{equation}
	where the fermions are coupled to HS bosons $\phi_n$ with no kinetic term: $\left< \phi_n (\tau) \phi_n (0) \right>_{\text{free}} = \delta (\tau)$. Averaging out the disorder in the replica-diagonal ensemble, we obtain the following action (See Appendix~\ref{app:action} for further details)
	\begin{subequations}
		\label{eq:allaction_1}
		\begin{align}
			\mathcal{S}   & =  \mathcal{S}_f  + \mathcal{S}_b  \,,\, \text{where } \label{eq:action_totalmodel} \\ 
			\mathcal{S}_f  & = \frac{N}2 \sum_{\omega_f} \left[-G(\omega_f) \Sigma(\omega_f) -  \ln (-\im \omega_f - \Sigma(\omega_f))   \right]   \label{eq:action_fmodel} \\
			\mathcal{S}_b  & = \frac1{2\beta} \sum_{n,\omega_b} \Big(  1- \lambda_n [G^2](\omega_b) \Big) |\phi_n(\omega_b)|^2 \label{eq:action_bmodel}  
		\end{align}
	\end{subequations}
	where $G$ and $\Sigma$ are the fermion propagator and self-energy, $\omega_{f/b}$ are fermion/boson Matsubara frequencies, and $[G^2](\omega_b)$ is $G(\tau)^2$ in frequency domain.
	
	In $S_b$, the boson mode $\phi_n(\omega_b)$ is governed by a quadratic potential, which cannot be unstable:
	\begin{equation}
		1-\lambda_n [G^2](\omega_b) \ge 0\,.
	\end{equation} 
	The mode $\phi_n(\omega_b)$ becomes condensed when equality is attained above, in the thermodynamic limit. Note that since $0 \le [G^2](\omega_b) \le [G^2](\omega_b = 0)$ for all $\omega_b$, only the modes with $\lambda_n = \lambda_{\max}$ and $\omega_b = 0$ can condense.
	
\begin{figure}
		\centering
		\includegraphics[width = 0.6\columnwidth]{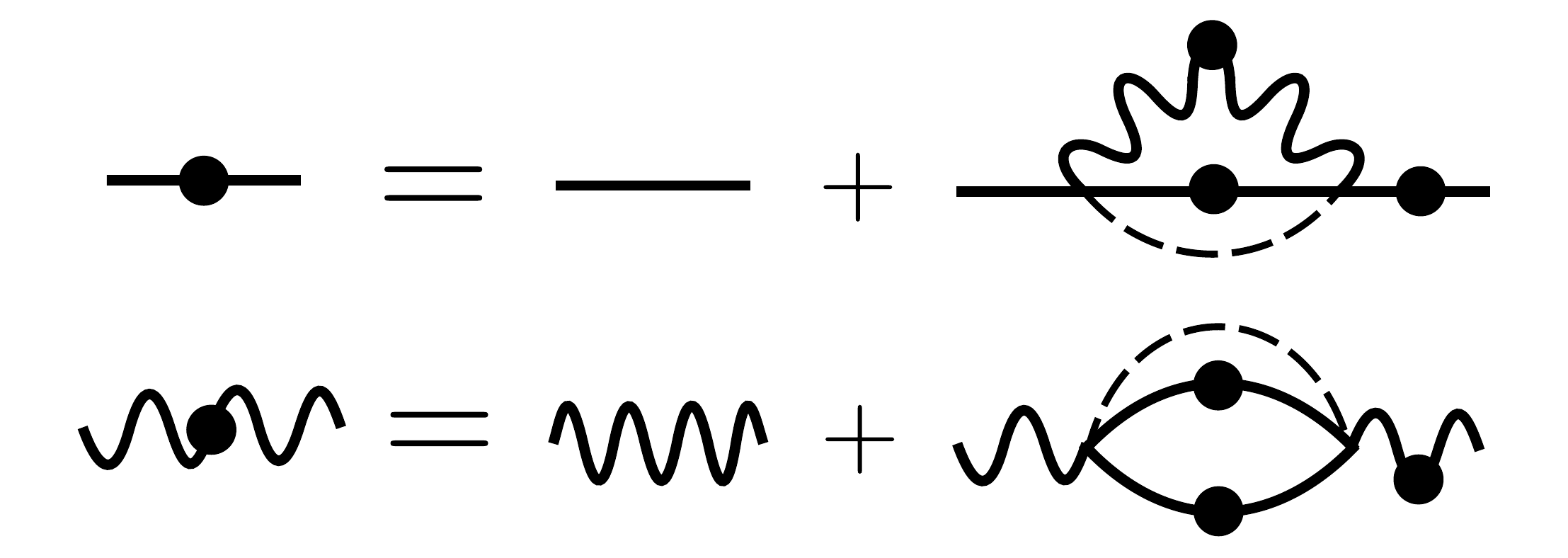}
		\caption{Diagrammatic representation of the Schwinger-Dyson equations~\eqref{eq:Gs} and \eqref{eq:Sigmas}. The fermion (boson) propagator is represented by a straight (wavy, resp.) line.  The dashed line denotes disorder contraction. A dot indicates a dressed propagator.}
		\label{fig:SDdiagram}
	\end{figure}

	To proceed further, we separate the condensed and normal modes:
	\begin{equation}
		\mathcal{S}_b = \mathcal{S}_{b,N} + \mathcal{S}_{b,C}  \,. 
	\end{equation}
	The normal modes are to be integrated out:
	\begin{equation}
		\mathcal{S}_{b,N} = \frac{N}{2\beta}  \int \sum'_{\omega_b} \ln(1-\lambda [G^2](\omega_b)) \rho(\lambda) d\lambda
	\end{equation}
	where the sum $\sum'$ excludes the condensed modes. The latter have macroscopic occupation and can be treated classically:
	\begin{align}
		\mathcal{S}_{b,C} &= N \left(1 - \lmax [G^2](\omega_b) \right) \Phi \,, \\
		\text{where }  \Phi  &:= \frac1{N \beta} \sum_{n: \lambda_n = \lmax} |\phi_n (\omega_b = 0)|^2  \label{eq:Phi}
	\end{align}
	In summary, we showed that the action $S$ is large-$N$, so the model can be solved by saddle-point Schwinger-Dyson (SD) equations:
	\begin{subequations}\label{eq:SDs}
		\begin{align}
			& G(\omega_f) = \frac1{-\im \omega_f - \Sigma (\omega_f)} \label{eq:Gs} \\
			&  G_\lambda (\omega_b) = \frac1{1 - \lambda [G^2](\omega_b)} \label{eq:Gboson} \\
			&  \Sigma (\tau) = 2 \gamma G(\tau) \int \lambda G_{\lambda}(\tau) \rho(\lambda) d\lambda + 2 \lmax \Phi G(\tau)  \label{eq:Sigmas} \\
			&  \Phi = 0 \text{ or } 1-\lmax [G^2](0)  = 0 \,. \label{eq:BEC}
		\end{align}
	\end{subequations}
	Above, we denote by $G_\lambda$ the propagator of normal bosons with $\lambda_n = \lambda$. The large-$N$ action and the SD equaions can be summarized by the Feynman diagrams in Fig.~\ref{fig:SDdiagram}, and will be the starting point of all subsequent analyses. 
	
	%
	

	\section{Non-Extensive Ranks}\label{sec:supersub}
	In this section, we review the cases where the rank $R$ is either much smaller or much larger than $N$. Although their physics are known from previous works, the analysis will provide useful insights to the study of the extensive rank regime.
	
	\subsection{Sub-Extensive Ranks}\label{sec:sub}
	Let us fist consider the regime of sub-extensive ranks, where the rank $R \ll N$, $\gamma \to 0$. In this regime, the non-trivial behavior of the model is completely determined by boson condensation. Indeed, the fermion self energy has only a condensate contribution, as \eqref{eq:Sigmas} reduces to  
	\begin{equation}
		\Sigma = 2 \lmax \Phi G \,. \label{eq:Sigma_sub}
	\end{equation}
	Consequently, the only way to obtain a nontrivial solution is to let $\lmax > 0$, which we shall assume in the rest of this subsection.
	
	Then, the trivial solution $\Phi = 0$, $ G(\tau) = \mathrm{sign}(\tau)/2$ is valid as long as $T > T_c := \lmax / 4$. At $T_c$, a boson condensation transition takes place. Below that, $\Phi >0$ and we have
	\begin{align}
		G(\omega_f) = \frac{2\im}{\omega_f + \sgn(\omega_f)\sqrt{8\lambda_{max} \Phi   + \omega_f^2}}  \,. \label{eq:G_free}
	\end{align}
	In turn, The value of $\Phi$ is determined by $\Phi > 0$ \eqref{eq:BEC} for any $T<T_c$. At low temperatures, $G(\tau)$ has a power-law decay with a SYK$_{2}$ (free fermion) exponent:
	\begin{equation}
		|G(\tau)| \sim \frac1{|\tau|^{2\Delta}} \quad \Delta = \frac12\,, \label{eq:Delta_freefermion}
	\end{equation}

	\subsection{Super-Extensive Ranks}\label{sec:super}
	Now let us consider the super-extensive rank regime. It is convenient to redefine how $R$ scales with $N$ as follows: 
	\begin{equation}
		R = \gamma N^\alpha, \ \alpha > 1, \ \gamma = \BigO(1)\,.
	\end{equation}
	The random couplings $u_{ij}^{(n)}$ should also be normalized differently:
	\begin{equation}
		\overline{ u_{ij}^{(n)} u_{kl}^{(m)} } = \frac1{N^{a}} \delta_{ik}\delta_{jl} \delta_{nm} \,,\, \ a = \frac{\alpha+3}2 >2 \,.
	\end{equation} 
	The last relation will turn out necessary and sufficient to ensure an extensive free energy for $\alpha > 1$.  
	Indeed, the fermionic action \eqref{eq:action_fmodel} is intact, and in the bosonic one \eqref{eq:action_bmodel}, $\lambda_n [G^2](\omega_b)$ is replaced by $ \lambda_n [G^2](\omega_b) N^{2-a} \ll 1$ at large-$N$ since $a > 2$. So, no condensation is possible. Moreover, we can expand the normal boson action:
	\begin{equation}
		\begin{split}
			\mathcal{S}_b = \mathcal{S}_{b,N} &= \frac1{2\beta} \sum_{n,\omega_b} \ln \Big(  1- \lambda_n N^{2-a}[G^2](\omega_b) \Big) \\
			&= -\frac12 \sum_{\ell=1}^{\infty}  \sum_{\omega_b, n}   \frac1{\ell} \lambda_n^\ell [G^2](\omega_b)^\ell N^{(2-a) \ell}
			\label{eq:SigmaFsuper}
		\end{split}
	\end{equation}
	and keep only the first \textit{non-trivial} term. That turns out to be $\ell = 2$, because the $\ell = 1$ term is a constant $E_0 \beta = -\beta N^{2-a} \sum_{n} \lambda_n / 8 $. Therefore, we have
	\begin{equation}
		\begin{split}
			\mathcal{S}_b  &= -\frac{N\beta}2 \, \mu_2   \int_\tau \frac14 G(\tau)^4 \,,\,  \mu_2 = 2\gamma\int \rho(\lambda) \lambda^2    d\lambda\,. \label{eq:Sbmu2}
		\end{split}
	\end{equation}
	This action is identical to that of the standard SYK$_{q=4}$ model~\cite{masaki16,Bi:2017yvx}.  
	Thus, at low temperatures, we have the well-known conformal solution~\cite{Maldacena:2016hyu}
	\begin{equation}
		|G(\tau)| \sim \frac{b}{|\tau|^{2\Delta}} \,,\, b = \frac{1}{\sqrt[4]{8\pi \mu_2 }} \,,\, \Delta = \frac14 \,. \label{eq:sykGsupersub}
	\end{equation}
	
	\section{Extensive Ranks}\label{sec:results}
	In the last section, we have shown that the low-rank SYK models reduce to SYK$_4$ at super-extensive ranks, and SYK$_2$ (or trivial) at sub-extensive ranks. To look for novel low-temperature behaviors, we shall focus on the regime of extensive rank, $R = \gamma N$, and resume the normalization of Sec.~\ref{sec:model}.
	
	
	\subsection{Crossover temperature}
	We start by determining the temperature regime where we expect new physics, as a function of rescaled rank $\gamma= R/N$.
	
	When $\gamma$ is large, we expect the model to reduce to SYK$_4$ in some temperature regime, by consistency with Sec.~\ref{sec:super} above. To find the crossover temperature, we apply the results there, extrapolated to the extensive regime $\alpha= 1$ and $a = 2$. The truncation of the Taylor series in \eqref{eq:SigmaFsuper} is valid if and only if $ |\lambda_n| [G^2](\omega_b) \ll 1$ for any $n$ and $\omega_b$. This is equivalent, assuming \eqref{eq:sykGsupersub}, to 
	\begin{equation}
		T \gg T_* = \sqrt{\mu_2} \exp\left[-\frac{\sqrt{2\pi \mu_2}}{\max_n |\lambda_n|}\right]
	\end{equation}
	where $\mu_2$ is defined in \eqref{eq:Sbmu2}. For a fixed distribution $\rho(\lambda)$,  $T_*$ depends on the rank in a stretched exponential fashion:
	\begin{equation}
		T_* = \sqrt{c_1 \gamma} e^{-\sqrt{c_2\gamma}} \label{eq:Tstar}
	\end{equation} 
	where $c_1$ and $c_2$ depend on $\rho(\lambda)$. 
	
	In summary, the model is governed by an unstable SYK$_4$ fixed point at intermediate temperatures $\sqrt{\mu_2}\gg T \gg T_*$. This transient regime exists only for large $\gamma$. For $\gamma \lesssim 1$, there is only one crossover at $T_* \sim \max_n |\lambda_n|$, from the trivial UV fixed point directly to a novel IR fixed point. The rest of the section will be devoted to characterizing the latter. 
	
	\subsection{Four-fold way}\label{sec:overview}
	The low-temperature behavior in the extensive-rank regime depends strongly on the shape of the distribution $\rho(\lambda)$. To prepare for a systematic study, we shall describe and motivate the classification table~\ref{table:ExtensiveCase}. 
	
	\begin{table*}
		\begin{center}
			\centering
			\begin{tabular}{|c||c|c|c|c|}
				\hline
				&\includegraphics[width=0.9in]{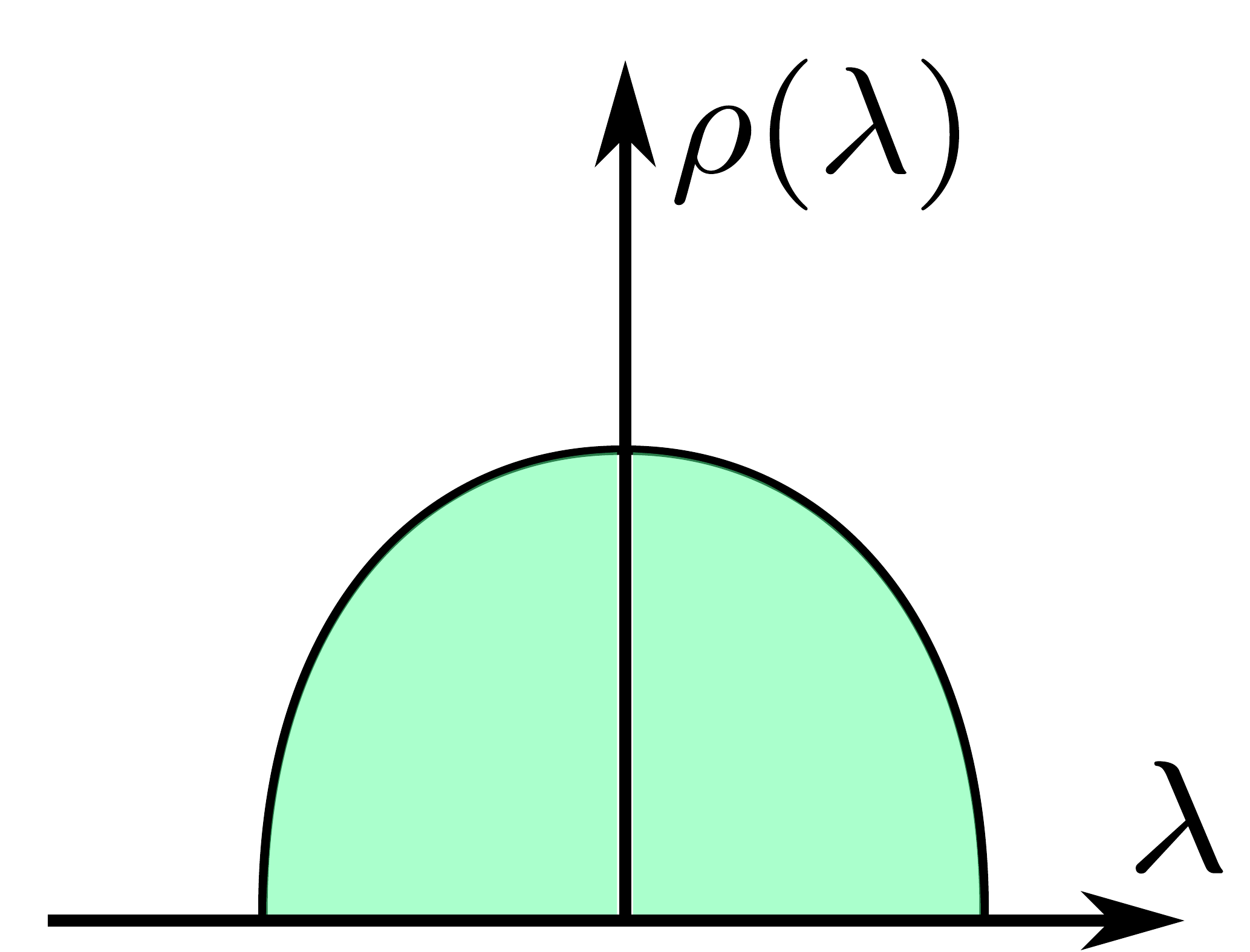}& 
				\includegraphics[width=0.9in]{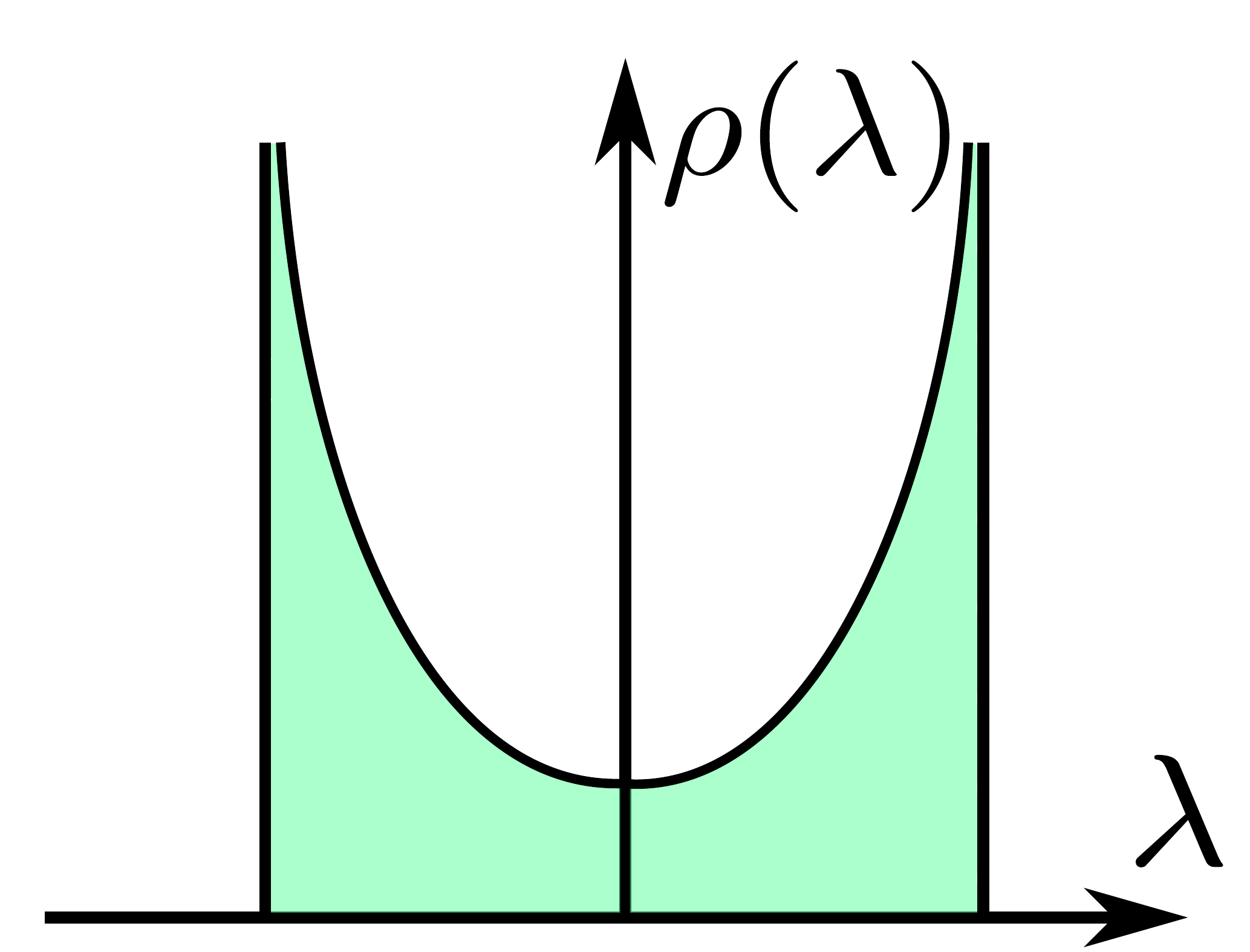}& 
				\includegraphics[width=0.9in]{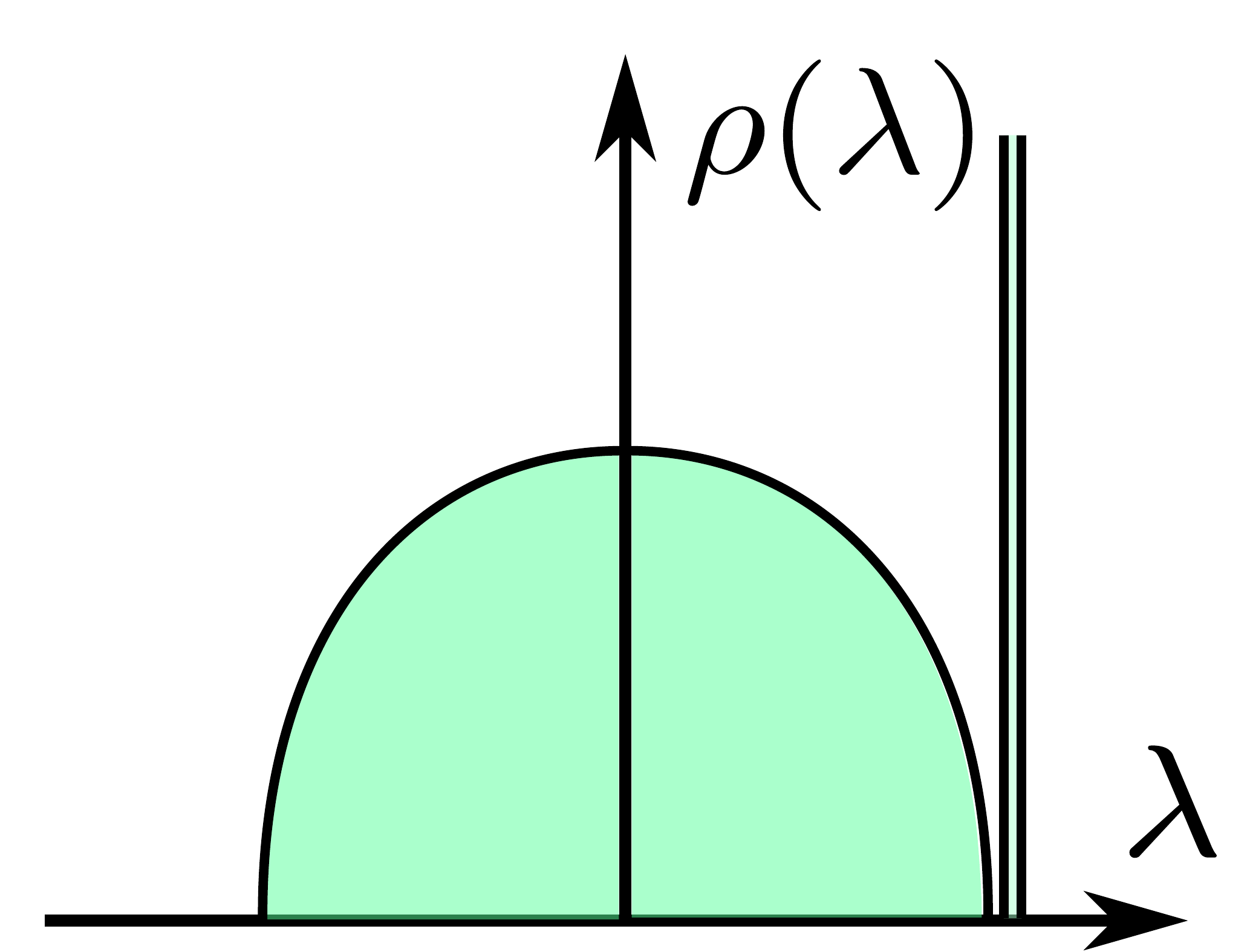}& 
				\includegraphics[width=0.9in]{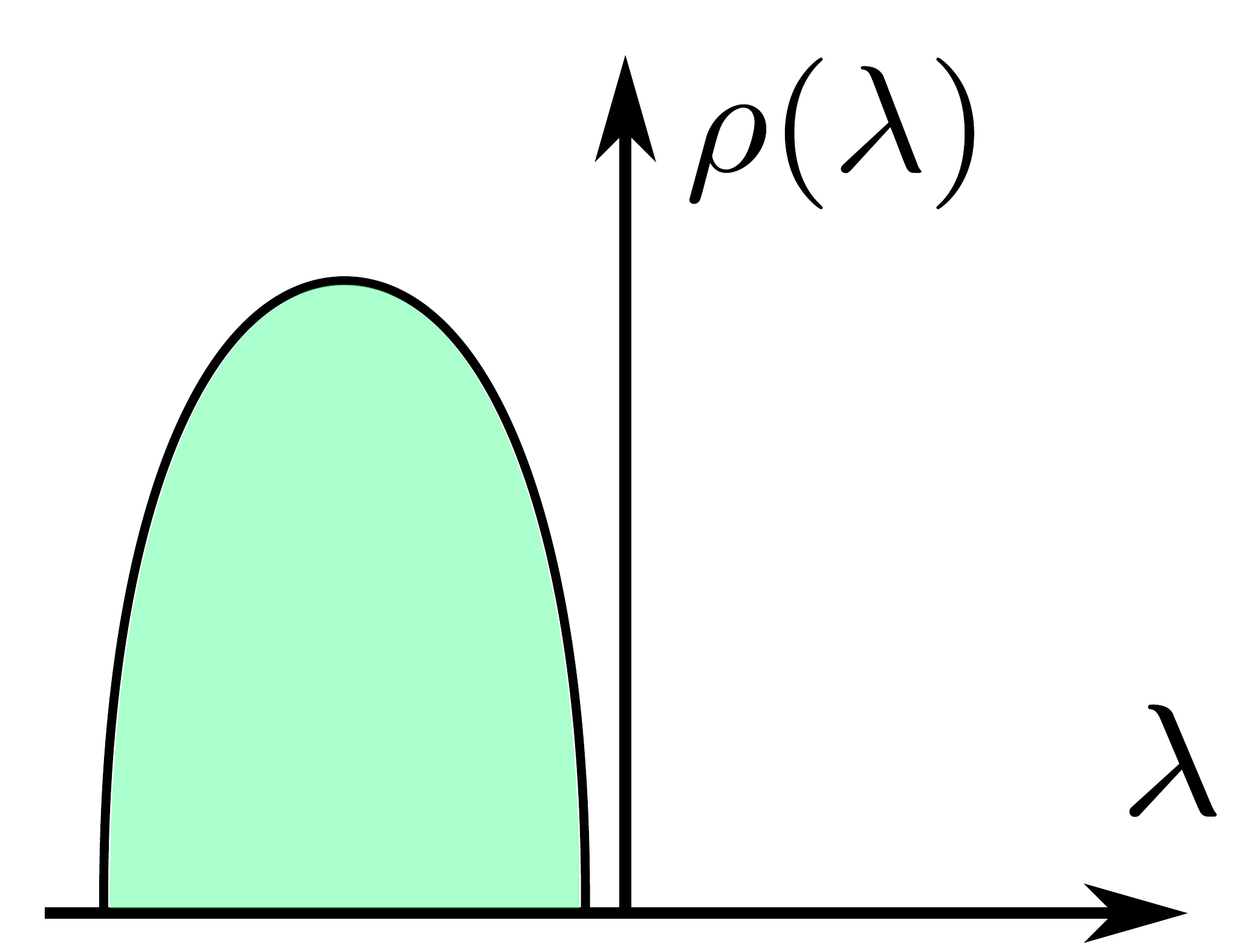} \\ \hline 
				Class & I & II & III & IV \\ \hline
				$|G(\tau)| \sim \tau^{-2\Delta}$ &$\Delta=1/2$ &  $ \Delta=1/2$ & $\Delta_\gamma\in (1/4, 1/2)$ & $\Delta_\gamma \in (0, 1/4)$ \\ \hline
				Broken $\mathcal{T}$? &  $T< T_c$ &  $T = 0$ & Never & Never \\ \hline
				$S$ (entropy) &$ c T $ & $ c T^{\nu} , 0 < \nu < 1 $ & $S_0 + cT $ &  $S_0 + c T $ \\ \hline 
				$\lambda_L$ (chaos) & $\sim T^{\eta+1}$ & $\sim T$, $\le 2 \pi T$ & $ 2\pi T $ &  $ 2\pi T $ \\ \hline 
			\end{tabular}
			\caption{
				A classification of different qualitative behaviors of the eigenvalue distribution $\rho(\lambda)$, and the resulting low-energy behaviors of the model at extensive rank. $\mathcal{T}$ is the time reversal symmetry. $\lambda_L$ is the quantum Lyapunov exponent.}
			\label{table:ExtensiveCase}
		\end{center}
	\end{table*}
	
	For that, let us recall the SD equation~\eqref{eq:Sigmas}. 
	The integral over boson modes on the RHS can be split into a condensate part ($\lambda = \lmax$, $\omega_b=0$, $G_\lambda = \infty$) and a normal part ($G_\lambda < \infty$), as follows: 
	\begin{align}
		&\Sigma(\tau) = 2 \gamma G(\tau) F(\tau) +   2 G(\tau) \lmax \Phi \label{eq:SigmaGF} \\
		& F(\omega_b) = f([G^2](\omega_b)) \,,\,  f(y) :=   \int \frac{\lambda \rho(\lambda) }{ 1 - \lambda y }  d\lambda  \,. \label{eq:f} 
	\end{align} 
	Above, $F$ is a weighted sum of the propagator of non-condensed bosons. It depends on $\rho(\lambda)$ via $f(y)$. The classification of $\rho(\lambda)$ will be based on the analytical properties of $f(y)$. 
	
	First, Class IV is defined by $\lmax \le 0$. Such distributions are clearly distinct from the rest in that $f(y)$ is analytical on the positive real axis $[0, +\infty)$. On the other hand, when $\lmax > 0$, $f(y)$ increases with $y$ and becomes maximal at the singularity at 
	\begin{equation}
		y_* := 1/\lmax \,.
	\end{equation} 
	The nature of the singularity is completely determined by the right edge of $\rho(\lambda)$ near $\lmax$. There are three possibilities/classes: 
	\begin{subequations}
		\begin{align}
			\text{I } & \lim_{y \to y_*} f(y) < +\infty \,, \label{eq:f(y)ClassI}  \\
			\text{II } & \lim_{y \to y_*} f(y) = +\infty \text{ but } f(y) \ll 1/(y_* - y) \,, \label{eq:f(y)ClassII} \\ 
			\text{III } & f(y) \sim c_0 /(y_* - y) \,,\, y\to y_* \,, c_0 \in (0,1] \,. \label{eq:f(y)ClassIII}
		\end{align}
	\end{subequations}
	In terms of the right edge of $\rho(\lambda)$, these classes are exemplified by the following (see Table~\ref{table:ExtensiveCase} for a cartoon):
	\begin{align*}
		\text{I } & \rho \sim (\lmax- \lambda)^{\eta} \,,\, \eta > 0 \text{ (vanishing edge)} \\
		\text{II } & \rho \sim (\lmax- \lambda)^{\eta} \,,\, -1 < \eta \le 0 \,, \text{ (non-vanishing edge)} \\ 
		\text{III } & \rho = c_0 \delta(\lambda - \lmax)+\dots \text{ (delta peak)}
	\end{align*}
	Note that, although the above example distributions do not exhaust all the possibilities (there can be log corrections to power laws), the classification in terms of $f$ is exhaustive. 
	
	Let us provide some further rationale for Class I-III. Class I is distinguished by $f(y\to y_*) < +\infty$, which is a necessary condition for condensation at finite $T$. Indeed, recall from \eqref{eq:BEC} that condensation requires $[G^2](\omega_b=0) = 1/\lmax = y_*$, and thus $F(\omega_b) = f(y_*)$ must be finite. Thus, finite-$T$ condensation only happens in Class I, and not in Classes II and III. Amongst the latter two, Class III is distinguished by a macroscopic degeneracy of the softest boson modes. This prevents condensation at even zero-$T$, making Class III rather resemble Class IV. In contrast, Class II is closer to Class I: as we shall see later, the softest boson modes do condense at zero-$T$.

	\subsection{Class III \& IV: SYK$_q$-like}\label{sec:Cl34}
	The low-temperature behavior of these classes have been partially studied in Refs.~\cite{Bi:2017yvx} and \cite{Esterlis:2019ola,wang19}, respectively. It was shown that the fermion Green's function $G(\tau)$ becomes conformal invariant at low temperatures:
	\begin{equation}
		G(\tau) = A\, \sgn(\tau) |\tau|^{-2\Delta} \,,\, 1\ll |\tau| \ll \beta \,, \label{eq:G_conformal}
	\end{equation}
	where $\Delta$ depends continuously on the rescaled rank $\gamma = R/N$. 
	
	Let us review how to find such a solution in Class IV. We claim that the SD equations have the following conformal approximations:
	\begin{subequations} \label{eq:repar_1}
		\begin{align}
			- G(\omega_f) \Sigma(\omega_f) = 1 \label{eq:repar_1G}  \\
			[G^2](\omega_b) F(\omega_b) = 1 \label{eq:repar_1F} \\
			\Sigma(\tau) = 2\gamma F(\tau)G(\tau) \label{eq:repar_1S} 
		\end{align}
	\end{subequations}
	Above, \eqref{eq:repar_1G} comes from $G(\omega_f)^{-1} = {-\im \omega_f - \Sigma (\omega_f)}$ \eqref{eq:Gs} by dropping $\im \omega_f$ at low frequencies, while \eqref{eq:repar_1F} assumes $[G^2](\omega_b) \to \infty$ as $|\omega_b| \to 0$ , which will be verified below, and which implies $$F(\omega_b) = f([G^2](\omega_b)) \sim \frac{1}{[G^2](\omega_b)} \int \rho(\lambda)d\lambda = \frac{1}{[G^2](\omega_b)}$$ by \eqref{eq:rholambda}.  \eqref{eq:repar_1S} is implied by \eqref{eq:SigmaGF} and the absence of condensation. Using standard Fourier transform formaulae, one finds that eqs.~\eqref{eq:repar_1} are satisfied by \eqref{eq:G_conformal} if the scaling dimension $\Delta$ satisfies
	\begin{equation}
		\gamma = \frac{(2 \Delta -1) (\sec (2 \pi  \Delta )-1)}{8 \Delta -2} \,,\, \Delta \in (0,1/4) \,, \label{eq:exponent}
	\end{equation}
	see Appendix~\ref{app:scaling} for details, and Fig.~\ref{fig:exponent_negative} for a numerical check. The fact that $\Delta < 1/4$ ensures the assumption behind \eqref{eq:repar_1F} above. In the limits $\gamma\to 0$ and $\gamma \to +\infty$, $\Delta$ tends to $0$ (the SYK$_{q\to\infty}$ value) and $1/4$ (the SYK$_4$ value), respectively. 
	
	A nice byproduct of the above analysis is that all boson propagators are equal in the scaling regime: 
	\begin{align}
		\lambda G_\lambda(\tau) = F(\tau)  \sim |\tau|^{-2\Delta_b} \label{eq:Glambda34}
	\end{align}
	where 
	\begin{align}
		\Delta_b = 1-2\Delta \, \label{eq:Deltab}
	\end{align}
	is the bosonic scaling dimension. Furthermore, the approximate SD equations \eqref{eq:repar_1} enjoy an reparametrization symmetry, just as those of SYK$_4$. Upon rewriting \eqref{eq:repar_1G} and \eqref{eq:repar_1F} in time domain, it is readily checked that any reparametrization $\tau \to f(\tau)$, transforms a solution \eqref{eq:repar_1} to another one in the following way:
	\begin{equation}
		\begin{split}
			& G(\tau,\tau') \rightarrow [f'(\tau)f'(\tau')]^{\Delta} G(f(\tau),f(\tau')) \\
			& F(\tau,\tau') \rightarrow [f'(\tau)f'(\tau')]^{\Delta_b} F(f(\tau),f(\tau')) \\
			& \Sigma(\tau,\tau') \rightarrow [f'(\tau)f'(\tau')]^{\Delta_b}\Sigma(\tau,\tau') \,.
			\label{eq:repar_3}
		\end{split}
	\end{equation}
	Adapting the argument of Ref.~\cite{Bagrets:2016cdf}, one may show that the reparametrization symmetry is broken explicitly and spontaneously, giving rise to a Schwartzian action of soft modes. A consequence of this broken symmetry~\cite{Maldacena:2016hyu,Bagrets:2016cdf,Kitaev:2017awl} is the maximal out-of-time order correlator growth (Lyapunov exponent), which we will show in Sec.~\ref{sec:chaos} below.
	
	\begin{figure}
		\centering
		\includegraphics[scale=.8]{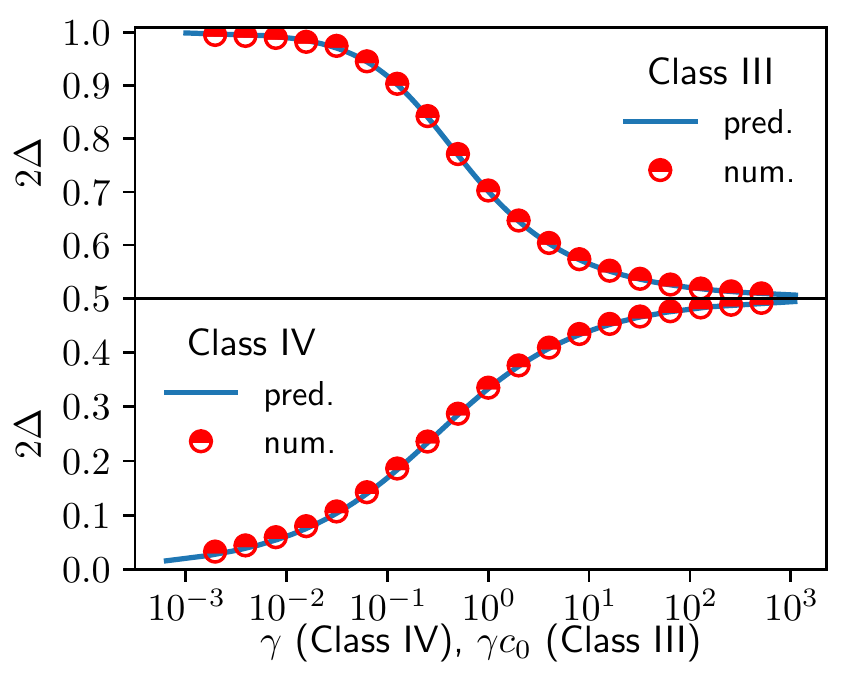}
		\caption{Fermion scaing dimension $\Delta$ as a function of the re-scaled rank $\gamma$ in Class III (top) and IV (bottom). The analytic curve is given by ~\eqref{eq:exponent} and \eqref{eq:exponent1} (with $c_0 = 1$), and compared to numerical solutions of the SD equations, with $\rho(\lambda) = \delta(\lambda \pm 1)$ for $\lmax \lessgtr 0$.}
		\label{fig:exponent_negative}
	\end{figure}
	
	The situation in Class III is formally similar, although physically different. Indeed, the equations \eqref{eq:repar_1G} and \eqref{eq:repar_1S} still hold (the latter does so because of no condensation), while \eqref{eq:repar_1F} becomes 
	\begin{equation}
		\big(1-\lmax[G^2](\omega_b)\big)F(\omega_b) = c_0\lmax \,, \label{eq:repar_FIII}
	\end{equation}
	according to \eqref{eq:f(y)ClassIII}. By a similar analysis, we find a conformal solution  such that $\lmax[G^2](\omega_b) = 1 - C |\omega_b|^{4\Delta - 1}$, where $\Delta$ is determined by $\gamma c_0$ as:
	\begin{equation}
		\gamma c_0 = \frac{(2 \Delta -1) (\sec (2 \pi  \Delta )-1)}{8 \Delta -2} \,,\, \Delta\in (1/4, 1/2) \,, \label{eq:exponent1}
	\end{equation}
	which we plot and test in Fig.~\ref{fig:exponent_negative}. Unlike Class IV, as $\gamma$ decreases to $0$, $\Delta \to 1/2$ increases to the SYK$_2$ value. Concerning the bosons, only the soft modes, with $\lambda_n = \lmax$, have a power-law propagator satisfying \eqref{eq:Glambda34} and \eqref{eq:Deltab}.

	\subsection{Class I and II: almost Fermi liquids}
	The analysis of Class I and II involves the shape of the distribution $\rho(\lambda)$ to a greater extent.  For simplicity, we shall focus on the following family of power-law edge singularities: $\rho(\lambda) \sim (\lmax - \lambda)^{\eta}$, so that
	\begin{equation}f(y) \sim 
		\begin{cases} (y_* - y)^{\eta} & \eta < 0  \\
			f_* - C (y_* - y)^{\eta} & \eta > 0 \,,
		\end{cases} \label{eq:f_powerlaw}
	\end{equation} 
	We shall assume $0<\eta < 1$ in Class I and $-1 < \eta < 0$ in Class II, although our analysis can be easily applied to other situations, e.g., a uniform distribution $\rho(\lambda) = \mathrm{const}$, which is a marginal case of Class II with $f(y) \sim -\ln (y_* - y)$. 
	
	In contrast to Class III and IV, the fermion scaling dimension is always 
	\begin{equation}
		\Delta = 1/2  \,.
	\end{equation}
	This can be seen by a simple argument: any Class I/II distribution can be approached from Class III, by taking a $c_0 \to 0$ limit. Then \eqref{eq:exponent1} implies $\Delta \to 1/2$. Physically, however, Class I and II are far from being the limit cases of Class III. Let's discuss them in turn. 
	
	For Class I, a boson condensation must form at low temperature. To see why it must be so, recall that the fermion self energy has two potential contributions, condensate and normal: 
	\begin{equation*}
		\Sigma(\tau) = \underbrace{2 \lmax {\Phi} G(\tau)}_{\Sigma_C} + \underbrace{2\gamma {F}(\tau)G(\tau)}_{\Sigma_N}  \text{ (Class I)} \,.
	\end{equation*}
	By definition of Class I, $F(\omega_b) = f([G^2](\omega_b)) \le y_*$ is bounded as $\omega_b \to 0$, so $F(\tau) \lesssim |\tau|^{-1}$, and $\Sigma_N(\tau) \ll |\tau|^{-2}.$ However, $\Delta = 1/2$ is equivalent to $\Sigma(\tau) \sim |\tau|^{-1}$, which must come from the condensate contribution 
	$\Sigma_C(\tau) \sim |\tau|^{-1} \,,$
	which is dominant. 
	
	Now, a consequence of the condensation is $[G^2](\omega_b = 0) = y_*$. Therefore, the sub-leading $\Sigma_N$ is affected by the singularity of $f$ at $y_*$ \eqref{eq:f_powerlaw}:
	\begin{align}
		F(\omega_b) &\sim |\omega_b|^{\eta} \,, \label{eq:F_ClassI} \\
		\Sigma_N (\omega_f) &\sim  -\im \sgn(\omega_f) |\omega_f|^{1+\eta} \, \label{eq:Sigma_N_ClassI}
	\end{align}
	when $T \ll |\omega_{b,f}| \ll 1$. This non-analytic sub-leading term in self-energy distinguishes Class I from a standard Fermi liquid, which has $\omega_f^2$ instead. One may view Class I as an ``almost Fermi liquid'' with an anomalously large quasi-particle decay rate. 

	The situation of Class II is similar, except for the following subtlety: condensation is impossible at finite temperature (see Sec.~\ref{sec:overview}). Nevertheless, as $[G^2](0) \nearrow y_*=1/\lmax$, $F(\omega_b = 0) = f([G^2](0))$ can be arbitrarily large. In particular, it can be $\propto \beta$ at low temperatures, and play the role of $\Phi$. Indeed, we can rewrite \eqref{eq:SigmaGF} as
	\begin{align}
		&\Sigma(\tau) = \underbrace{2 \lmax \widehat{\Phi} G(\tau)}_{\Sigma_C} + \underbrace{2\gamma \widehat{F}(\tau)G(\tau)}_{\Sigma_N} \,,\, \text{ where}  \label{eq:Sigma_renorm_main}  \\
		& \widehat{\Phi} :=  \frac{\gamma  \overline{F}}{\lmax} \,,\, 
		\widehat{F}(\tau) := F(\tau) - \overline{F} \,,\, \overline{F} := \frac{ F(\omega_b = 0) }{\beta}\,. \nonumber
	\end{align}
	Above, we redefined $\Sigma_C$ (and $\Sigma_N$) in terms of the effective condensate $\widehat{\Phi}$. A similar argument as above shows that, at low temperature, $\widehat{\Phi}$ remains positive. This means that 
	\begin{equation}
		y_* - [G^2](\omega_b= 0)  \sim T^{-1/\eta} {\ll} T  \,,\,  T\to 0  \label{eq:yminusG20}
	\end{equation}
	(since $0 > \eta > -1$) is negligibly small, so that \eqref{eq:F_ClassI} and \eqref{eq:Sigma_N_ClassI} still hold (of course, the meaning of $\Sigma_N$ and the value of $\eta$ are different). So, like Class I, Class II realizes an almost Fermi liquid, with a higher quasi-particle decay rate at low frequency. The physical consequences of this will be studied below. 
	
	To close, let us discuss the spontaneous breaking of time reversal symmetry $\mathcal{T}$, which is closely related to soft boson modes. Indeed, while the Hamiltonian is even under $\mathcal{T}$, the fermion bilinears $Q_n$~\eqref{eq:Qndef} are odd. Therefore,  a condensed boson mode generates a term $\phi_n Q_n$, which breaks $\mathcal{T}$.
	It follows immediately that $\mathcal{T}$ is broken at low temperatures in Class I (and also in the sub-extensive regime). In Class II, although no condensation takes place at finite temperature, $\mathcal{T}$ is broken at zero temperature. To see this, note that the softest boson mode with $\lambda_n = \lmax$ has the following propagator:
	\begin{equation*}
		G_{\lmax}(\omega_b = 0) = T^{1/\eta} \gg 1/T \,,
	\end{equation*}
	according to \eqref{eq:yminusG20}. This means that the following order parameter diverges
	$$
	\frac1\beta \int_0^\beta \left< \phi_n(\tau) \phi_n(0) \right> d\tau \to \infty
	$$
	as $T\to0$, which implies the breaking of $\mathcal{T}$ symmetry at zero temperature~\cite{Bi:2017yvx}. Repeating the analysis for Class III and IV,  uing the results in Sec.~\ref{sec:Cl34}, it is not hard to show that $\mathcal{T}$ is unbroken even at zero temperature in both SYK$_q$ classes.

	\section{Thermodynamics}\label{sec:thermodynamics}
	In this section, we study the low temperature thermodynamics of the four classes of the extensive-rank regime both analytically and numerically. The free energy of the model is given by the saddle-point action $F = \mathcal{S}_{\text{saddle}} / \beta$, where $S$ is as defined in \eqref{eq:action_totalmodel}. From that, it is not hard to obtain the energy density:
	\begin{equation}
		\begin{split}
			-\frac{\beta E}{N} &=   \frac{1}2 \beta \Phi +
			\frac12 \gamma  \sum_{\omega_b} [G^2](\omega_b) F(\omega_b)  \label{eq:energy_therm}
		\end{split}
	\end{equation}
	We shall study the low-temperature thermodynamics in both sub-extensive and extensive regimes, by a combination of analytical and numerical methods. 

	\subsection{Sub-extensive ranks}
	The low-temperature thermodynamics can be calculated exactly in the sub-extensive regime. The only contribution to the energy is the condensate:
	\begin{equation}
		\varepsilon := E/N = -\frac12 \Phi\,, \label{eq:epsilon_sub}
	\end{equation}
	which is determined by \eqref{eq:BEC} and \eqref{eq:G_free}, rewritten as:
	\begin{align}
		&T \sum_{k = 0}^{\infty} g_\Phi(\pi T + 2 \pi k T) = 1 \,,   \label{eq:sum_therm} \\
		&\text{where } g_{\Phi}(\omega_f) :=  \frac{8 \lmax}{\left(\omega_f + \sqrt{8\lmax \Phi + \omega_f^2}\right)^2}\,.
	\end{align}
	For small $T$, the sum can be estimated with the Euler-McLaurin formula, 
	\begin{align}
		1
		= &  \int_{\pi T}^{\infty} \frac{d\omega_f }{2\pi} g_{\Phi}(\omega_f) + \frac{T}2 g_\Phi(\pi T)  
		- \frac{\pi T^2}{6}  g'_\Phi(\pi T) + \cdots \nonumber   \\
		= & \int_{0}^{\infty} \frac{d\omega_f }{2\pi} g_\Phi(\omega_f) 
		+ \frac{\pi T^2}{12}  g'_\Phi(\pi T) + \cdots \label{eq:smallT_sub0therm}  \\
		= & \int_{0}^{\infty} \frac{d\omega_f }{2\pi} g_\Phi(\omega_f) 
		+ \frac{\pi T^2}{12}  g'_\Phi(0) + \cdots \,. \label{eq:smallT_subtherm}
	\end{align}
	Above, we denoted $g' := \partial_{\omega_f} g$; in the second line, we approximated the integral $\int_0^{\pi T} g$ by exanding $g$ at $\omega_f = \pi T$; throughout, the omitted terms $\in \BigO(T^3)$. Equating the first term in \eqref{eq:smallT_subtherm} to $1$, and evaluating some integrals, we obtain 
	\begin{align}
		\Phi &=  \Phi_0 + c_V T^2+ \BigO(T^3) \text{ where} \\
		\Phi_0 &:= \frac{8}{9\pi^2} \lmax  \,,\,  c_V := \frac{\pi^2}{8 \lmax}  \,.
	\end{align}
	Consequently, by \eqref{eq:epsilon_sub}, the specific heat 
	\begin{equation}
		C_V = c_V T + \BigO(T^2) \,.
	\end{equation}  
	is linear in $T$. Note that, only the numerical value of $\Phi_0$ and $c_V$ depend on the exact form of $g$, while $C_V \propto T$ only depends on the fact that $\partial_{\omega_f} g$ and $\partial_\Phi g$ both exist, are continuous and nonzero whenever $\Phi > 0$. 
	
	\subsection{Class I and II}
	We now extend the above exact analysis to Classes I and II, by by making some approximations. As the method is similar for both Classes,  let us explain it just for Class I in some detail. 
	
	To start, we observe that the SD equations \eqref{eq:Gs} and \eqref{eq:SigmaGF} imply that
	\begin{align}
		&G(\omega_f) = \frac{2 \im}{J +  \mathrm{sign}(\omega_f) \sqrt{ 8 \lmax \Phi + J^2} } \label{eq:iteration_main} \\
		&   \text{where }  J := \omega_f -  \im \Sigma_{N}(\omega_f) \,.
	\end{align}
	To make progress, we make two approximations. First, by \eqref{eq:Sigma_N_ClassI}, $J \approx \omega_f$ is independent of $\Phi$ at low frequencies. 
	\begin{enumerate}
		\item We ignore the $\Phi$ dependence of $J$, and approximate it by its leading small $\omega_f$ behavior. In Class I, we have $J \sim \omega_f$ by \eqref{eq:Sigma_N_ClassI}.
		\item  We approximate the energy by $ \varepsilon\approx -\frac12 {\Phi},$ ignoring the $\propto \gamma$ terms in \eqref{eq:energy_therm}.
	\end{enumerate}
	These approximations renders the problem nearly  identical to the sub-extensive case. Indeed, $\varepsilon \approx -\Phi/2$ is determined by the same equation \eqref{eq:sum_therm} with the same $g_{{\Phi}}$. Therefore, we predict that 
	\begin{equation}
		C_V \propto T  \text{ (Class I)} \,, \label{eq:C_V_ClassI}
	\end{equation}
	at least for small $\gamma$.
	
	We now apply the same approximations to Class II, while switching $\Phi$ for $\widehat{\Phi}$ everywhere. Now, notice that since $\eta < 0$ in Class II, $J \sim |\omega_f|^{1+\eta}$ by \eqref{eq:Sigma_N_ClassI}. Thus, $g_{\widehat{\Phi}}$ is non-analytical in $\omega_f$ at $\omega_f = 0$: the derivative $g'_{\widehat{\Phi}}(\omega_f) \sim |\omega_f|^{\eta}$ is divergent as $\omega_f \to 0$, so that \eqref{eq:smallT_sub0therm} now implies
	\begin{equation}
		1 - \int_{0}^{\infty} \frac{d\omega_f }{2\pi} g_{\widehat{\Phi}} (\omega_f) \sim T^2  \,
		g'_{\widehat{\Phi}}(\pi T) \sim T^{2 + \eta}\,. \label{eq:T_twopluseta_therm}
	\end{equation}
	Consequently, we predict that the specific heat is anomalously large at low-$T$:
	\begin{equation}
		C_V \sim T^{1 + \eta } \text{ (Class II)} \,.  \label{eq:C_V_ClassII}
	\end{equation}

	\subsection{Numerical Results}
\begin{figure}
		\centering
		\includegraphics[width=.8\columnwidth]{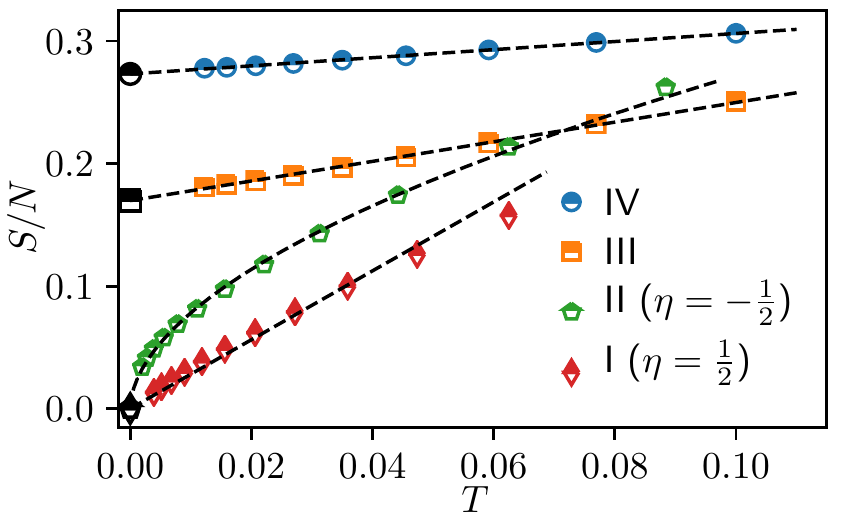}
		\caption{Entropy density $S/N$ as function of the temperature $T$ in the four universality Classes. The data points are obtained from numerical solutions of the SD equations, and are well fitted by(dashed curves): $S = cT$ for I, $S = c T^{\nu}$ ($\nu = 0.5(4)$) for II, and $S = S_0 + c T$ for III and IV. The black markers are the extrapolated zero-$T$ entropy. For display, the entropy for Class II is multiplied by $1.5$. See \cite{supp}, Sec. F for further details.}
		\label{fig:entropy}
	\end{figure}

	We now compute the temperature dependence of entropy in all four classes in the extensive rank regime, by solving the large-$N$ SD equations numerically. Representative results are given in Fig.~\ref{fig:entropy}. 
	
	In Classes III and IV, the data is well described by
	\begin{equation}
		S/N = S_0 + c_V T + \cdots \,,
	\end{equation}
	where the zero-temperature entropy is positive $S_0 > 0$. This nonvanishing zero-temperature entropy imply that the Class III and IV models are reminiscent of the SYK$_q$ model.
	
\begin{figure}
		\centering
		\includegraphics[width=.8\columnwidth]{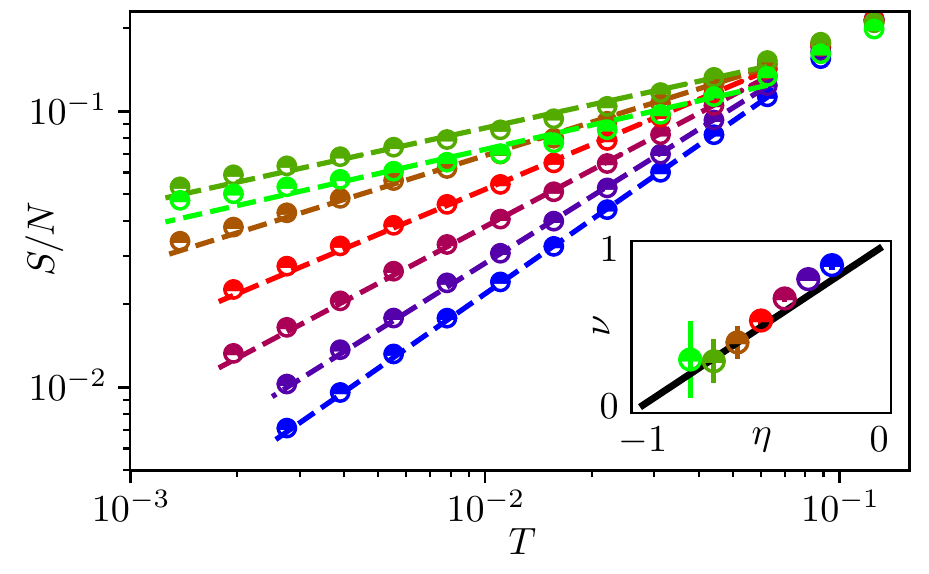}
		\caption{Numerical test of the prediction~\eqref{eq:C_V_ClassII} for Class II.    Main plot: Entropy density $S/N$ as function of temperature $T$, with $f(y) = y (1-y)^{\eta}$ for $\eta = -0.8, -0.7,  \dots, -0.2$ (top to bottom) , and $\gamma = 0.2$ (except that $\gamma = 0.1$ for $\eta = -.8$). The dots are from numerical solution of the SD equation. The dashed lines are best fits to a power law $S / N = c T^\nu$. Inset: the fit exponent $\nu$ (dots, same color code as main plot), compared to the prediction~\eqref{eq:C_V_ClassII} (solid line). }
		\label{fig:entropy_1}
	\end{figure}

	In stark contrast, we find that neither Class I nor II has an extensive residual entropy, and the entropy obeys a power law
	\begin{equation}
		S/N \stackrel{T\to 0}\propto \begin{cases} T & \text{Class I} \\ 
			T^{\nu}\,,\, 0< \nu < 1 & \text{Class II}\,,
		\end{cases}
	\end{equation}
	which are consistent with the predictions \eqref{eq:C_V_ClassI} and \eqref{eq:C_V_ClassII} above, since $C_V = T \partial S/\partial T$. We  computed the exponent more thoroughly, albeit for relatively small ranks, and found a good quantitative agreement with \eqref{eq:C_V_ClassII}, see Fig.~\ref{fig:entropy_1}. 
	
	\section{Out-of-time order correlator}\label{sec:chaos}
\begin{figure}
		\centering
		\includegraphics[width=.8\columnwidth]{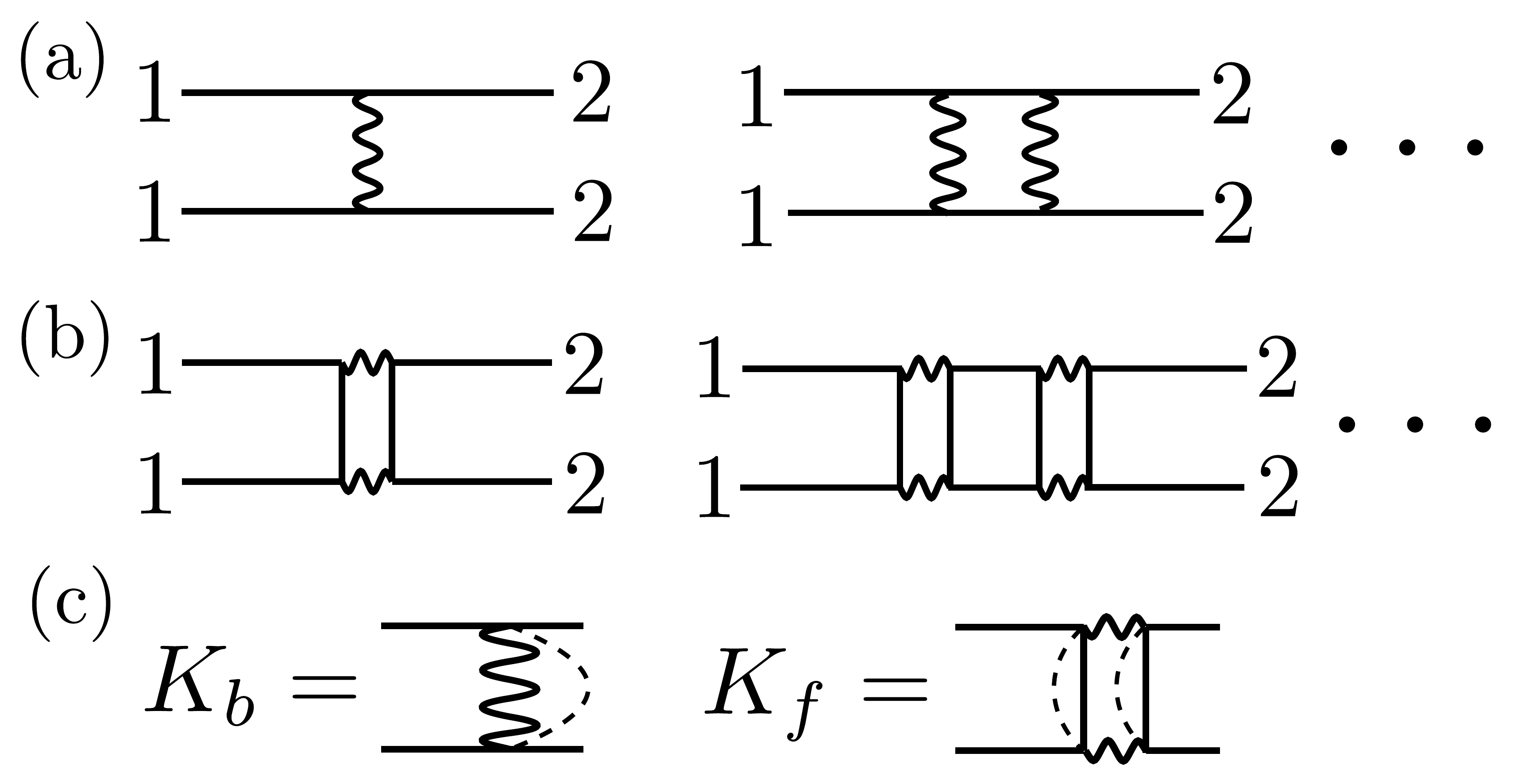}
		\caption{(a,b) Examples of ladder diagrams contributing to the out-of-time order correlator~\eqref{eq:OTOC}. Disorder lines are omitted for display.  (c) The kernels generating the ladders, with disorder lines. All propagators are dressed. }
		\label{fig:kernels}
	\end{figure}

	We now study the growth of the out-of-time order correlator (OTOC):
	\begin{equation}
		\overline{\mathrm{Tr}\left[y\gamma_1(t_1) y\gamma_1(0) y\gamma_2(t_2) y\gamma_2(0) \right]}\,,\, y = \frac{e^{-\beta H / 4}}{\mathrm{Tr}(e^{-\beta H} )}  \,. \label{eq:OTOC}
	\end{equation}
	Following closely the approach of Refs~\cite{Kitaev:2015,Maldacena:2016hyu,Banerjee:2016ncu}, we focus on the $\BigO(1/N)$  and  exponentially growing  part of the OTOC, given by the sum of a series of ladder diagrams generated by two types of ladder rungs. The ladder kernel is $K = K_b +  K_f$, where (see Fig.~\ref{fig:kernels}):
	\begin{subequations}\label{eq:Ks}
		\begin{align}
			K_b(t_{1, \dots, 4}) &= \frac{2}{N} \sum_{n}  G_R(t_{13})G_R(t_{24})G_{\lambda_n, lr}(t_{34})\label{eq:Kb} \\
			K_f(t_{1, \dots, 4}) &= \frac{4}{N} \sum_{n} \int  dt_5 dt_6 \ G_R(t_{15})G_R(t_{26}) \times \label{eq:Kf}   \\
			&\lambda^2 G_{\lambda_n, R}(t_{35}) G_{\lambda_n, R} (t_{46})G_{lr}(t_{34})G_{lr}(t_{56})  \,.\nonumber 
		\end{align}
	\end{subequations}
	Above, $t_{ij} := t_i - t_j$, the subscript ``$R$'' indicates a retarded propagator, and ``$lr$'' a Wightman correlator~\cite{Maldacena:2016hyu}; both can be obtained from the Euclidean-time correlator. 
	
	We then compute the quantum Lyapunov exponent $\lambda_L$ by finding an eigenfunction 
	\begin{equation}
		\int d t_1 d t_2  K (t_{1,\dots,4}) \mathcal{F}(t_1,t_2) = k \mathcal{F}(t_3,t_4)   \label{eq:eigen}
	\end{equation}
	of the form $\mathcal{F}(t_1,t_2)= f_F(t_{12}) e^{\frac{\lambda_L}2 (t_1 + t_2)}$ and with eigenvalue $k = 1$~\cite{Kitaev:2015,Maldacena:2016hyu,Banerjee:2016ncu}.
	
	\subsection{Class III \& IV: maximal chaos}
	We now carry out the calculation described above in the conformal limit of Class III and IV. This can be done analytically, thanks to the results of Sec.~\ref{sec:Cl34} above (see also Appendix~\ref{app:scaling}). 
	
	Let us first discuss Class IV, where both fermionic and bosonic propagators are conformal:
	\begin{align}
		G(\tau) &= A \, \sgn{\tau} |\tau|^{-2\Delta}\,, \quad 1 \ll |\tau| \ll \beta \\
		\lambda G_{\lambda}(\tau) &= F(\tau) = \frac{(1-4\Delta)}{2\pi A^2 \tan 2\pi\Delta} |\tau|^{2-4\Delta}
		\label{eq:GandF_chaos}
	\end{align}
	The constant $A$ will drop out in the final results. The retarded and wightman correlators are obtained by analytical continuations to real time~\cite{Maldacena:2016hyu}:
	\begin{align}
		G_R(t) &= 2A\cos(\pi\Delta) \theta(t) \left[\frac{\pi}{\beta \sinh\frac{\pi t}{\beta}} \right]^{2\Delta} \\
		G_{lr}(t) &= A\left[\frac{\pi}{\beta \cosh\frac{\pi t}{\beta}} \right]^{2\Delta}  \,, \label{eq:G_Rlr}
	\end{align}
	and similarly for the bosons. Therefore, the summed terms in \eqref{eq:Ks} are independent of $n$, so the sum $\frac1N \sum_n$ can be simply replaced with $R /N= \gamma$: 
	\begin{equation}
		\begin{split}
			K_b(t_{1, \dots, 4}) &= 2 \gamma  G_R(t_{13})G_R(t_{24})F_{lr}(t_{34})  \\
			K_f(t_{1, \dots, 4}) & =  4\gamma\int dt_5 dt_6 G_R(t_{15})G_R(t_{26}) \\
			& F_R(t_{53})F_R(t_{64})G_{lr}(t_{34})G_{lr}(t_{56}) \,. \label{eq:kernel_GF}
		\end{split}
	\end{equation}
	The RHS of the above equations involve only known conformal propagators, and will be analyzed exactly. 
	
	Before doing so, we argue that \eqref{eq:kernel_GF} holds for Class III as well, provided we replace $\gamma \to c_0 \gamma$ [note that $\Delta$ is also a function of $c_0\gamma$ instead of $ \gamma$, see \eqref{eq:exponent1}]. This is because the sum over bosons in \eqref{eq:Ks} are dominated by the softest ones, with $\lambda_n = \lmax$. There are $c_0 \gamma N$ of those, and their propagator still satisfies \eqref{eq:GandF_chaos}. 
	
	We now look for eigenfunctions of $K = K_f + K_b$ with the following Ansatz~\cite{Kitaev:2015,Maldacena:2016hyu}:
	\begin{equation}
		\mathcal{F}(t_1, t_2) = e^{-h\frac{\pi}{\beta}(t_1+t_2)} \left[\frac\pi{\cosh\frac{\pi}{\beta}t_{12}}\right]^{2\Delta-h} \,,
		\label{eq:Fansatz}
	\end{equation}
	where the Lyapunov exponent is related to $h$ by $\lambda_L = -2h \pi T$. By a straightforward but tedious calculation (going back and forth between the time and frequency domains), we can show that ${\mathcal{F}}$ is indeed an eigenfunction of both $K_b$ and $K_f$, with the following eigenvalues:
	
	\begin{align}
		k_b(h) &= \frac{(1-2 \Delta ) \sin (2 \pi  \Delta ) \Gamma (1-2 \Delta )^2 \Gamma (2 \Delta -h)}{\pi  \Gamma (-h-2 \Delta +2)} \,,\, \nonumber \\
		k_f(h) &= \frac{2 \left(8 \Delta ^2-6 \Delta +1\right) \sin (2 \pi  \Delta ) \sin (4 \pi  \Delta )}{\pi ^2 \Gamma (-h-2 \Delta +2) \Gamma (4 \Delta -h)} \times \label{eq:totalkernel} \\
		&  \Gamma (1-2 \Delta )^2 \Gamma (4 \Delta -1)^2 \Gamma (-h-4 \Delta +2) \Gamma (2 \Delta -h) \,.  \nonumber 
	\end{align}
	The eigenvalue of the total kernel $k(h) := k_f(h) + k_b(h)$ has a remarkable property: for any $\Delta \in (0, 1/2)$, $k(h) = 1$ if and only if $h = -1$. As a consequence, the low rank SYK model in the extensive regime with Class III or IV distributions is maximally chaotic:
	\begin{equation}
		\lambda_L = 2\pi T \quad \text{Class III, IV, $T \ll \max_n |\lambda_n|$.}
	\end{equation}
	\begin{figure}
		\centering
		\includegraphics[width=0.8\columnwidth]{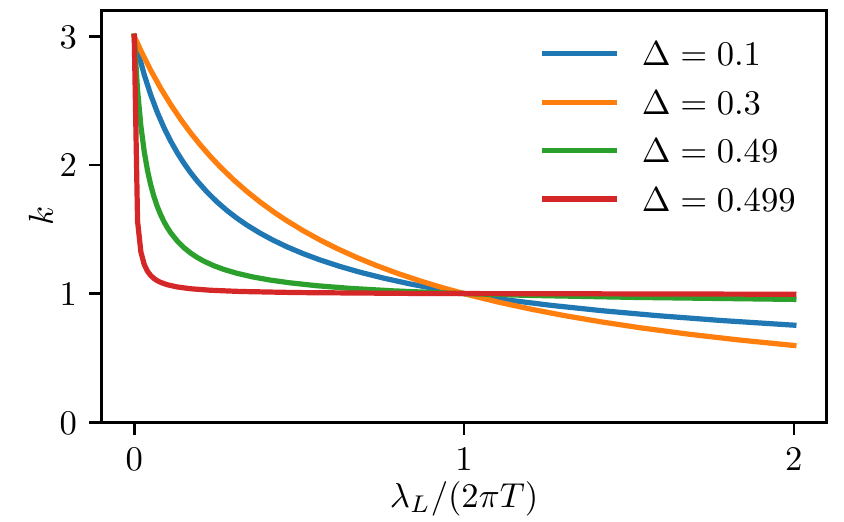}
		\caption{The eigenvalue $k$ of the ladder kernel corresponding to \eqref{eq:Fansatz}. The analytical expression is given in (S.55). }
		\label{fig:Kh}
	\end{figure}

	\subsection{Class I \& II: non-maximal chaos}
	We now briefly discuss the cases of Class I and II. The key difference of these classes is that, determining $\lambda_L$  requires going beyond the fermion scaling dimension $\Delta$. Indeed, the kernel eigenvalues \eqref{eq:totalkernel} satisfy
	\begin{equation}
		\lim_{\Delta\to1/2} k_b(h) = 1 \,,\, \lim_{\Delta\to1/2} k_f(h) = 0 \,, 
	\end{equation}
	for any $h$. So $k(h) = k_b(h) + k_f(h)$ also tends to $1$ in that limit (this can be seen in Fig.~\ref{fig:kernels}), and $\lambda_L$ cannot be determined by the above method. This situation also occurs in the Fermi-liquid phase of Ref.~\cite{Banerjee:2016ncu}, and with the SYK$_q$ model in the $q \to 2$ limit. In all these cases, $\lambda_L$ depends on the sub-leading terms in the propagators.
	
	A detailed analysis along this line, which we will present in an upcoming work, leads to the following results. In Class I, The $T$-dependence of $\lambda_L$ is reminiscent of the $\omega$ dependence of the quasi-particle decay rate \eqref{eq:Sigma_N_ClassI}:
	\begin{equation}
		\lambda_L \sim T^{1 + \eta} \,,\, 0 < \eta < 1 \,. \label{eq:chaos_Class1}
	\end{equation}
	Therefore,   Class I is more chaotic than a Fermi liquid where $\lambda_L \propto T^2$~\cite{Banerjee:2016ncu,guo19}. Naively extrapolating \eqref{eq:chaos_Class1} to Class II, we would have a violation of the bound on chaos $\lambda_L \le 2\pi T$. Yet, a more careful analysis indicates that $\lambda_L \propto T$, but the bound is not always saturated by the pre-factor.

	\section{An application}\label{sec:experiments}
	Recently, Ref.~\cite{2018PhRvL.121c6403C} proposed an interesting realization of the SYK model in a graphene flake with irregular boundaries, using quantum Hall ferromagnetism. A strong magnetic flux $\Phi$ is induced onto the graphene flake, creating $\Phi/\Phi_0$ degenerate lowest Landau levels (LL$_0$) in the presence of chiral symmetry. Here, $\Phi_0$ is a flux quanta $hc/e$. Since the graphene boundary is irregular, the LL$_0$ wave-functions are pseudo-random. Hence, Projecting the Coulomb interaction onto the LL$_0$ then produces a disordered four-fermion interaction, which the authors of Ref.~\cite{2018PhRvL.121c6403C} claimed to be of SYK$_4$ nature. 
	
	
	Now, we argue that the realized Hamiltonian more likely a low rank SYK, in the extensive-rank regime and of Class IV. For this, let $\vert \varphi_j \rangle$ be the LL$_0$ wave-functions, and $c_j$ be the associated  fermionic annihilation operator. Then the projected Coulomb interaction is
	\begin{equation*}
		H = \sum_{ijkl}
		\sum_{r,r'} V(r-r') \braket{r|\varphi_i} \braket{\varphi_j|r}  c_i^\dagger c_j
		\braket{r|\varphi_k} \braket{\varphi_l|r}  c_k^\dagger c_l \,.
	\end{equation*}
	Above, $r$ and $r'$ runs over all the lattice sites, and $V(r-r')$ is the Coulomb potential, which we can diagonalize as
	\begin{equation}
		V(r-r') = - \sum_{n} \lambda_n U_{rn} U_{r'n} \,,
	\end{equation} 
	where $U_{rn} = \pbraket{n\vert r} \in \R$ forms a real orthogonal matrix, and $\lambda_n$ are the eigenvalues. Therefore,
	\begin{align}
		H &= -\sum_n \lambda_n Q_n^2 \,,  \\ &\textrm{ where }
		Q_n = \sum_{ij}  {u_{ij}}  c_i^\dagger c_j \,,\, 
		u_{ij} = \sum_r \braket{n\vert r} 
		\braket{r|\varphi_i} \braket{\varphi_j|r} \nonumber
	\end{align}
	Note that $Q_n$ is a Hermitian fermion bilinear. At this point, if we approximate $Q_n$ by a set of independent random fermion bilinears, we will have the complex version of the low rank SYK model, with an extensive rank.  Finally, the repulsive nature of Coulomb interactions implies that $\lambda_n \ge 0$ for all $n$, resulting in a Class IV distribution. 
	
	In summary, our argument reveals an additional structure in the seemingly random four-body interaction. It will be interesting to study whether there exists further relevant structures. If there are none and the realized model is indeed a Class IV, extensive-rank SYK, the goal of Ref.~\cite{2018PhRvL.121c6403C} will be still fulfilled. Indeed, as we showed above, Class IV is a maximal chaotic scrambler almost indistinguishable from SYK$_q$ for some $q > 4$.
	
	\section{Discussion}\label{sec:discussion}
	We have introduced and solved the low-rank SYK models, unifying and completing previous results~\cite{masaki16,Bi:2017yvx,Esterlis:2019ola,wang19}. The four classes of quantum phases that the model possesses, summarized in Table~\ref{table:ExtensiveCase}, fall into two categories. The fast scramblers of Class III and IV are equivalent to SYK$_q$ in all aspects we have studied, although the reparametrization symmetry in Class III is worth further elaborating. On the other hand, the almost Fermi liquids of Class I and II may not have reparametrization symmetry. However, they are stable under weak quadratic perturbations (since such a term is already generated dynamically). 
	
	The fermion-boson coupling form~\eqref{eq:H} of our model generalizes the electron-phonon coupling model of Refs~\cite{Esterlis:2019ola,wang19} in the normal state (\cite{supp}, Sec. H). These authors considered a Class III distribution of couplings $\rho(\lambda) = \delta(\lambda - \lmax)$. We showed that a non-degenerate distribution will belong to Class I or II (Class IV is impossible in this setting since $\lambda_n$ is always positive), which is almost a Fermi liquid. It will be interesting to understand the instability of such a phase into the superconducting state.
	
	Finally, our model in the extensive regime restores the physical rank of the coupling matrix in $SU(M)$ random quantum magnets away from the large $M$ limit. Our results thus suggest that the critical low-energy state of the magnet at finite $M$ is almost a Fermi liquid, probably of Class I, which contains the semi-circle law. Yet, by engineering a coupling matrix with a Class II-IV spectrum, one can still realize faster scramblers in atom-cavity settings. 
	
	\begin{acknowledgments}
		We thank Aavishkar Patel, Ionut-Dragos Potirniche and Thomas Scaffidi for helpful discussions and collaboration on related projects. We acknowledge support from the ERC synergy Grant UQUAM (EA and XC) and DOE grant DE-SC0019380 (EA and XC).
	\end{acknowledgments}

	\begin{appendix}
		\begin{widetext}
			\subsection{Large $N$ Action and Schwinger-Dyson Equations}\label{app:action}
			In this section we derive the action of the low rank SYK model. We first focus on the ``replica diagonal ensemble'' given by the disorder averaged partition function $\overline{Z^1}$ at inverse temperature $\beta$. Before we start, however, we will relax \eqref{eq:Jijkl} and \eqref{eq:uij_coupling} in order to also discuss the sub-extensive and super-extensive rank regimes. We modify \eqref{eq:Jijkl} and \eqref{eq:uij_coupling} to
			\begin{equation}
				R = \gamma N^\alpha + \textrm{sub-leading corrections}, \qquad \qquad
				\overline{ u_{ij}^{(n)} u_{kl}^{(m)} } = \frac1{N^a} \delta_{ik}\delta_{jl} \delta_{nm} \,.
				\label{eq:uij_coupling_app}
			\end{equation}
			$\alpha \in [0,2]$, and $\gamma$ is an order unity constant. Note that the parameter $a$ controls the normalization of the Hamiltonian. Requiring extensive energy fluctuation at \textit{infinite} $T$, we can find a relation between $a$ and $\alpha$:
			\begin{align}
				& \overline{\Tr[H^2] - \Tr[H]^2} \sim  N^{4-2a+\alpha} \,. \label{eq:fluc}
			\end{align}
			The fluctuation scales extensively with $N$ provided
			\begin{equation}
				a = (\alpha + 3)/2 \,.
				\label{eq:InfTempNorm}
			\end{equation}
			In particular, we have  $a = 3/2$ for a finite rank interaction $\alpha = 0$; For a near full rank interaction $\alpha = 2$, $a = 5/2$. For the extensive scaling in the main text, we have $\alpha = 1$,  $a = 2$. In general, however, normalization of the Hamiltonian at infinite $T$ may be different from that at finite $T$. As we will come to later, for sub-extensive ranks $a = 2$ in order to have an extensive free energy at finite temperatures.
			
			Having a rough idea of the normalization, let us get back to the large $N$ action.
			\begin{equation}
				\overline{Z} = \int [\mathcal{D}\gamma] \, \overline{e^{-\int_{\tau} d\tau \mathcal{L}}} \,,\, 
				\mathcal{L} = \sum_j  \gamma_j \dot{\gamma_j} + H \,. \label{eq:partitionfunction}
			\end{equation}
			As mentioned in the main text, after a Hubbard-Stratonovich (HS) decoupling, the Lagrangian is given as the following:
			\begin{align}
				& \mathcal{L} =  \sum_j  \gamma_j \dot{\gamma_j} + \sum_n \left( \lambda_n^{\frac12} \phi_n Q_n + \frac{\phi_n^2}2 \right) \,. 
				\label{eq:HSLagrangian}
			\end{align}
			Then, averaging over disorder results in the bi-local effective action
			\begin{equation}
				\begin{split}
					& S =  \int_\tau \left(  \sum_j  \gamma_j \dot{\gamma_j} + \sum_n \frac12{\phi_n^2} \right) - \frac12 \int_{\tau,\tau'} \sum_{nij} N^{-a} \lambda_n (\phi_n  \im \gamma_i \gamma_j)(\tau) 
					(\phi_n \im \gamma_i \gamma_j)(\tau')  \,.
				\end{split} \label{eq:S_BF_disorder}
			\end{equation}
			
			We now introduce as usual the Green function $G(\tau, \tau') = \frac1N \sum_j \gamma_j (\tau) \gamma_j(\tau')$ and impose the relation by adding the lagrange multiplier $${N} \Sigma (\tau, \tau') \left(G(\tau, \tau') - \sum_{j} \gamma_j(\tau)\gamma_j(\tau')\right)$$ to the action, where $\Sigma$ is the self-energy. Integrating out the fermions results in large-$N$ actions in the main text.

			\subsection{Details on the Scaling Analysis of Class III \& IV}\label{app:scaling}
			In this section we derive \eqref{eq:exponent} and \eqref{eq:exponent1}. The main tool is the following Fourier transform formulae:
			\begin{align}
				& \int e^{\im \tau \omega} |\tau|^{-a} \mathrm{sign}(\tau) d\tau = -2 \im \cos \left(\frac{\pi  a}{2}\right) \Gamma (1-a) \mathrm{sign}(\omega) \left| \omega \right| ^{a-1}\,, \ \ \int e^{\im \tau \omega} |\tau|^{-a}  d\tau \nonumber = 2 \sin \left(\frac{\pi  a}{2}\right) \Gamma (1-a) \left| \omega\right|^{a-1}
			\end{align}
			It is important to notice that, when we apply the above formulae to some $g(\tau)$ that is described by a power law only for large $\tau$, $g(\omega)$ will be given by the RHS plus a constant that depends on the UV details.
			
			Let us look for the conformal solution
			$$ G(\tau) \sim A \mathrm{sign}(\tau)\tau^{-2\Delta}$$
			that is compatible with the SD equations with appropriate approximations that make everything a power law. In all cases, we make the standard approximation $G(\omega) = -1/\Sigma(\omega)$. Note that it is crucial to keep the pre-factors (the power-laws alone do not constrain $\Delta$). For \eqref{eq:exponent}, we also approximate $f(y)$ to be $y^{-1}$. Straightforward computations yield
			\begin{align*}
				& G(\omega) \sim 2\im A\Gamma(1-2\Delta)\cos(\pi\Delta)\omega^{2\Delta-1} \\
				& [G^2](\omega) \sim 2A^2\Gamma(1-4\Delta)\sin(2\pi\Delta)\omega^{4\Delta-1}  \\
				& \Sigma(\omega) \sim -\frac{2\im \gamma\cot(2\pi\Delta)\cos(\pi\Delta)}{A \pi}\frac{\Gamma(2-4\Delta)\Gamma(2\Delta-1)}{\Gamma(1-4\Delta)} \omega^{1-2\Delta}
			\end{align*} 
			at low frequency or long time. Imposing $G(\omega)\Sigma(\omega) = -1$ gives \eqref{eq:exponent}; the condition $\Delta < 1/4$ ensures $ [G^2](\omega) \to 0$ as $\omega \to 0$, justifying the approximation of $f(y)$ by $y^{-1}$.
			
			The case of \eqref{eq:exponent1} is similar. $f(y)$ is approximated by $c_0 (y_* - y)^{-1}$ where $y_* = 1/\lmax$ is the nearest positive singularity of $f$. To apply this approximation, we look for solutions such that $\Phi = 0$ (no condensate) and that $[G^2](\omega) \to y_*$ as $\omega \to 0$ (this constant value depends on the UV details of $G$); then $ y_* - [G^2](\omega)$ is a power-law that only depends on the IR limit of $G$. With this in mind, the actual computation is almost the same as for \eqref{eq:exponent} above. The condition $\Delta > 1/4$ ensures that $[G^2](\omega) - [G^2](0) \sim |\omega|^{4\Delta -1}$ is vanishing. 
			
			We provide some details on Fig.~\ref{fig:exponent_negative} in the main text. For each data point, we numerically solve the SD equations for $\beta \in [10^2, 10^3]$ and extract $\Delta$ as follows: for each $\beta$, we compute the minimum of the log derivative $\Delta_{\beta} = -\min_{\tau} [d (\ln G)/ d (\ln \tau)]$, and then extrapolate to $\beta \to \infty$ using the Ansatz $\Delta_{\beta} = \Delta + a / \beta + b / \beta^c$. The errors are comparable to the marker size.
			
			\subsection{A Related Boson-Fermion Model}\label{app:boson}
			In this section, we consider a variant of the low-rank SYK model, which allows us to make connection with Ref.~\cite{Esterlis:2019ola,wang19}. As aforementioned, the four-fermion interactions of low-rank SYK model can be equivalently mediated by interactions with ``boson modes'' that do not have a kinetic term see ~\eqref{eq:H}. We now consider the effect of modifying the action by making the free boson action more ``realistic'':
			\begin{equation}
				\frac12 \int d \tau \phi_n(\tau)^2 \to \frac12 \int d \tau \phi_n(\tau) \left[ m^2 - \partial_\tau^2 \right] \phi_n(\tau)  \,,
			\end{equation}
			where $m > 0$. We shall focus on the extensive rank regime. 
			
			Following Appendix~\ref{app:action}, one can show that only the bosonic action~\eqref{eq:action_bmodel} is altered:
			\begin{subequations}
				\begin{align}
					S_b   & = \frac12 \sum_{n,\omega_b} \Big(\omega_b^2 + m^2 - \lambda_n [G^2](\omega_b) \Big) |\phi_n(\omega_b)|^2 \,,
					\label{eq:appendix_action_dynamic}
				\end{align}
			\end{subequations}
			Integrating out the non-condensed bosons and adding the condensate contribution leads to
			\begin{align}
				S_b = &\frac{N}2  \gamma \int  \rho(\lambda) \sum_{\omega_b} 
				\ln ( m^2  + \omega_b^2 - \lambda [G^2](\omega_b) ) d\lambda + \frac{N\beta}2 \Phi  (m^2 - [G^2](0) \lmax)  \,, \label{eq:Scomplete_app}
			\end{align}
			where the condensate fraction $\Phi$ is still defined by ~\eqref{eq:Phi} as only the zero-frequency modes can condense; $\Phi > 0$ if $\lambda_{\max} [G^2](0)  = m^2$. Among the Schwinger-Dyson equations, only the one involving the summed boson propagator $F$ is changed:
			\begin{equation}
				F(\omega_b) = \int \frac{ \rho(\lambda) \lambda }{m^2 + \omega_b^2 - \lambda [G^2](\omega_b)} d \lambda \label{eq:F_app}
			\end{equation}
			Although the relation between $F(\omega_b)$ and $[G^2](\omega_b)$ can no longer be encoded in a function $f(y)$, the quantum critical behavior found in the main text, summarized in Table~\ref{table:ExtensiveCase}, will remain essentially intact. This is because in any case, the low-frequency singularity of $[G^2](\omega_b)$ has a power law  $\sim |\omega_b|^{4\Delta - 1} \gg  \omega_b^2$ (as $\Delta < 1/2$), so we can drop the term $\omega_b^2$ in \eqref{eq:F_app} without affecting the low-frequency behavior. Then it is not hard to check that in Classes IV and III, the critical exponent $\Delta$ is still governed by \eqref{eq:exponent} and \eqref{eq:exponent1}, respectively, whereas $\Delta =1/2$ in Classes I and II: the whole low-rank perturbative theory carries through. 
			
			On the other hand, the super-extensive rank case needs more care. Restoring the $N^{2-a}$ factors in \eqref{eq:F_app} and exanding around $\lambda = 0$ gives (Although $N$ is originally the system size, it is more appropriate here to view it as a finite large parameter with which we take the high-rank limit from the extensive rank regime.)
			\begin{align}
				F(\omega_b) &= F_1(\omega_b) + F_2(\omega_b) + \dots \nonumber \\
				&= \frac{\mu_1 N^{2-a}}{m^2 + \omega_b^2} + \frac{\mu_2  N^{2(2-a)}}{(m^2 + \omega_b^2)^2} [G^2](\omega_b) + \dots \label{eq:F_app_expand}
			\end{align}
			where $a > 2$ and $\mu_\ell = \int \rho(\lambda) \lambda^\ell d \lambda$. The self energy has a similar expansion:
			\begin{equation}
				\Sigma(\tau) = \sum_{\ell=1}^\infty \Sigma_\ell(\tau) = \sum_{\ell=1}^\infty  2\gamma N^{\alpha-1} F_\ell(\tau) G(\tau) \label{eq:sigma_expand_app}
			\end{equation} 
			where $\alpha > 1$. Again, we want to determine the relation between $\alpha$ and $a$ to ensure the correct thermodynamics when $N \to \infty$. 
			
			Unlike in Section \ref{sec:super}, the term $\ell = 1$ can no longer be ignored, and the $\ell = 2$ term is not exactly $q=4$ SYK anymore. However, those do not affect the low temperature limit~\cite{Esterlis:2019ola,wang19}. Indeed, the extra factor $1/(m^2+\omega_b^2)$ in $F_2$ does not change the low-frequency behavior of $[G^2](\omega_b)$. For the $\ell = 1$ term,  \eqref{eq:sigma_expand_app} and \eqref{eq:F_app_expand} implies $$\Sigma_1(\tau) \le \gamma F_1 (\tau) =  N^{1-a + \alpha} \gamma \mu_1 \frac{e^{-|\tau|m}}{2m} $$
			decays exponentially. Therefore, if we adopt the scaling $\alpha = 2a - 3$, then the $\ell = 1$ term will become subdominant when $\tau m \gg \frac{(\alpha - 1)}{2} \ln N$. Meanwhile, at intermediate temperature, the model is dominated by the $\ell = 1$ term; this is reminiscent of the (un-stable) ``impurity'' fixed point in Ref.~\cite{Esterlis:2019ola}.
		\end{widetext}
	\end{appendix}
	
\end{document}